\documentclass[%
reprint,
superscriptaddress,
 amsmath,amssymb,
 aps,
pra,
]{revtex4-2}
\usepackage{graphicx}
\usepackage{dcolumn}
\usepackage{bm}
\usepackage{comment}
\usepackage{braket}
\usepackage{float}
\usepackage[caption=false]{subfig}
\newcolumntype{P}[1]{>{\centering\arraybackslash}p{#1}}
\newcolumntype{M}[1]{>{\centering\arraybackslash}m{#1}}
\graphicspath{{Figures/}}

\begin{document}

\preprint{APS/123-QED}

\title{Controllable entangled state distribution in a dual-rail reconfigurable optical network
}

\author{Shuto Osawa}
\email[e-mail: ]{sosawa@bu.edu} \affiliation{Dept. of Electrical and Computer Engineering \& Photonics Center, Boston University, 8 Saint Mary's St., Boston, MA
02215, USA}
\author{David S. Simon}
\email[e-mail: ]{simond@bu.edu} \affiliation{Dept. of Physics and Astronomy, Stonehill College, 320 Washington Street, Easton, MA 02357} \affiliation{Dept. of
Electrical and Computer Engineering \& Photonics Center, Boston University, 8 Saint Mary's St., Boston, MA 02215, USA}
\author{Vladimir S. Malinovsky}
\affiliation{US Army Research Laboratory, Adelphi, MD 20783, USA}
\author{Alexander V. Sergienko}
\email[e-mail: ]{alexserg@bu.edu} \affiliation{Dept. of Electrical and Computer Engineering \& Photonics Center, Boston University, 8 Saint Mary's St., Boston,
MA 02215, USA} \affiliation{Dept. of Physics, Boston University, 590 Commonwealth Ave., Boston, MA 02215, USA}

\date{\today}

\begin{abstract}
Reconfigurable distribution of entangled states is essential for operation of quantum networks connecting multiple devices such as quantum memories and quantum computers. We introduce new quantum distribution network architecture enabling control of the entangled state propagation direction using linear-optical devices and phase shifters and offering reconfigurable connections between multiple quantum nodes.
The basic two-photon entanglement distribution scheme is first introduced to illustrate the principle of operation. The scheme is then extended to a network structure with increased number of spatial modes connecting potential end-users. We present several examples of controllable network configuration modifications using time-dependent phase shifters.
\end{abstract}

\maketitle



\section{Introduction}
A significant amount of effort has been devoted to developing principles of quantum communication and quantum networking.
Controllable transport and distribution of quantum entangled states between multiple quantum devices connected in the network, such as quantum memories and quantum computers, is as important as the quantum state preparation or detection.
Such networks could provide a critical infrastructure for building a future quantum internet \cite{kimble2008quantum,pant2019routing}.
Much work has been devoted lately to demonstrating physical principles that enable entanglement of two distant quantum state preparation devices such as cold atoms \cite{duan2001long,duan2004scalable,zhao2009long,nicolas2014quantum,stephenson2020high} or Nitrogen-vacancy centers in diamond \cite{englund2010deterministic,humphreys2018deterministic,kalb2017entanglement}.

Long-distance photon entanglement naturally provides a link between distant quantum nodes. 
One of the major current challenges is how to scale the point-to-point elementary quantum-entangled links into a network allowing one to perform quantum operations between multiple users at will and without losing entanglement quality.
Transportation of quantum states has been studied in the context of perfect state transfer (PST) \cite{christandl2004perfect,chapman2016experimental,xu2020frequency,perez2013coherent}, and quantum state routing \cite{hahn2019quantum,zhan2014perfect,sazim2015retrieving}.

The Hong-Ou-Mandel (HOM) effect \cite{hong1987measurement} is a two photon quantum interference effect and often appears in quantum metrology applications \cite{dowling2008quantum,motes2015linear}.
The general design of the elementary dual-rail photon link enabling manipulation of correlated two-photon states, including a new higher-dimensional HOM effect, has been recently demonstrated \cite{osawa2020higher}.
The main goal of this paper is to establish operation and design principles for a reconfigurable entangled state distribution network of Bell states between numerous users.
The scheme is scalable and able to support many spatial modes and end-users.
The design presented here non-invasively selects propagation directions for amplitudes in the network by applying only phase shifts in the  system. The controllable time-dependent phase shift technology in waveguided structures is readily available. 
Electrical/thermal phase control is employed in multiple phase-encoding quantum key distribution \cite{minder2019experimental,bunandar2018metropolitan}. 
Phase control using embedded microheaters is actively used in large-scale photonic quantum information processing devices \cite{bhaskar2020experimental,larsen2019deterministic,asavanant2019generation,wang2017experimental,peruzzo2014variational,van2010integrated,masood2013comparison}.
Boson Sampling, which relies on phase control, has been experimentally implemented on multiple occasions and provided technological availability of on-chip waveguided photonic devices \cite{wang2019boson,tillmann2013experimental,bentivegna2015experimental}.

Mach-Zehnder (MZ) interferometers are the main resource for such network-based implementation schemes.
When MZ interferometers are used to control and route photons, they can be utilized only in a feed forward manner. The photon must be switched to go through at least a single additional closed loop every time it needs to reverse the direction of propagation. This means one must install a loop and a switch, engaging the return, at every node of a fully reconfigurable network. The approach in the manuscript saves considerably in terms of physical resources, and results in faster average transmission times due to the availability of shorter, loop-free routes for the photons to travel to any destination on the complex network.

In addition, the system has increased stability to phase shifts. Random phase shifts that affect adjacent input/output edges equally (for example due to slight bending of the system or an ambient temperature drift) affect the two entangled photons equally and therefore have no effect on the outcome. In the dual-rail system, only phase shift that differ significantly from one line to the next have significant effect.

\vspace{0pt}
\begin{figure}[htp]
\centering
\subfloat[]{\includegraphics[clip,width=0.4\columnwidth]{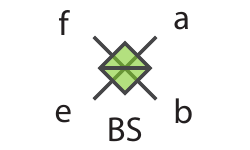}\label{}}
\subfloat[]{\includegraphics[clip,width=0.4\columnwidth]{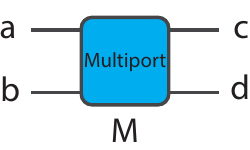}\label{}}
\caption{
Optical components used as linear transformers. (a) A beam splitter labeled as BS. The mathematical description of the device is given in Eq. (\ref{eqn:BS_transform}). (b) A four-port devices labeled as M. The mathematical operation follows Eq. (\ref{eqn:multiport_trans}).
}
\label{fig:basic_components}
\end{figure}

Two major linear-optical elements have been used as building blocks in the new design to effectively manipulate quantum probability amplitudes between different spatial modes.
One is a traditional beam splitter (BS) enabling equal splitting of amplitudes for quantum fields propagating in one direction and establishing (2x2) correspondence between two spatial modes (see Fig. \ref{fig:basic_components} (a)):
\begin{eqnarray}
\label{eqn:BS}
&\begin{pmatrix}
\hat{a}
\\
\hat{b}
\end{pmatrix}=
\frac{1}{\sqrt{2}}
\begin{pmatrix}
1 & 1\\
-1 & 1
\end{pmatrix}
\begin{pmatrix}
\hat{e}\\
\hat{f}
 \end{pmatrix},
\end{eqnarray}
where $\hat{e}$, $\hat{f}$ describe input photons, and $\hat{a}$, $\hat{b}$ describe output photons leaving after the BS transformation.
We drop the hat and dagger notation after introducing a symbolic notation describing BS transformation of photon amplitudes in mode $e$ and $f$,
\begin{eqnarray}
e \xrightarrow{BS} \frac{1}{\sqrt{2}}(a-b)\; \mbox{ and }\;f \xrightarrow{BS} \frac{1}{\sqrt{2}}(a+b),
\label{eqn:BS_transform}
\end{eqnarray}
The second element is the recently introduced \cite{simon2020quantum,osawa2020higher} totally reversible linear-optical scattering element realizing the Grover matrix correspondence (4x4) between four spatial modes (see Fig. \ref{fig:basic_components} (b)).
Experimental demonstration of bulk optics-based reversible three-ports as well as optical tritter based reversible three-ports have been demonstrated \cite{osawa2018experimental,kim2021implementation}.
The four-port version needs to be implemented on a chip for better stability.
The increase in matrix representation dimensionality of this four-port device is a reflection of the directionally-unbiased character of such linear-optical operations - any port could serve as an input and output at the same time, in contrast to a common beam splitter:
\begin{equation}
\label{eqn:Grover}
Grover =
\frac{1}{2}
\begin{pmatrix}
-1&1&1&1\\
1&-1&1&1\\
1&1&-1&1\\
1&1&1&-1
\end{pmatrix} ,
\end{equation}
where the rows and columns represent the four input and output modes.

In the same way as for the BS transformation, photon amplitudes in modes $a$ and $b$ are transformed by the Grover matrix in the following manner,
\begin{eqnarray}
a &\xrightarrow{M}& \frac{1}{2}(-a+b+c+d)\; \mbox{ and }\;\nonumber \\
b &\xrightarrow{M}& \frac{1}{2}(a-b+c+d),
\label{eqn:multiport_trans}
\end{eqnarray}
where M is a short notation for the action of the multiport device realizing the Grover matrix operation. Photons entering ports $c$ and $d$ are transformed in a similar manner.

\subsection{Manipulation of correlated two-photon states}

\begin{figure}[tbh]
	\centering
	\includegraphics[clip,width=2in]{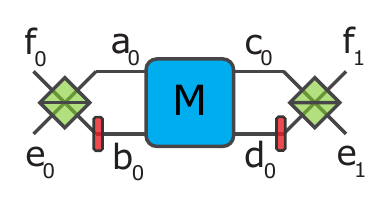}
	\caption{The basic unit used to control the flow of photons through the network is the multiport-based control unit (MCU) consisting of a Grover four-port, two beam splitters, and possibly phase shifters (the red rectangles). It is assumed that the values of the phase shifts can be switched between $0$ and $\pi$. When the phase is zero, the corresponding phase shifters will be omitted from the diagrams.
	}
	\label{mcufig}
\end{figure}

\begin{figure}[tbh]
\vspace{-7mm}
	\centering
	\includegraphics[clip,width=\columnwidth]{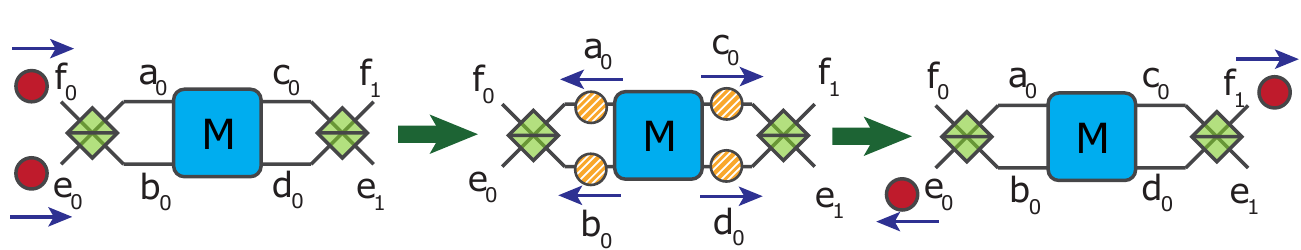}
	\caption{Manipulation of indistinguishable correlated photons. Two correlated photons enter the first BS from the left. The photons are transformed by BS followed by multiport M, resulting in a superposition state of right-moving and left-moving amplitudes distributed between two rails (shaded orange circles). The total photon number remains two. Finally, the right moving and left moving amplitudes get transformed by beamsplitters again, producing the original state but distributed between the right side and the left side of the multiport unit.
	}
	\label{fig:review_operation}
\end{figure}

We first briefly review the principles of correlated two-photon input state manipulation using Grover four-ports and beam splitters combined in the configuration illustrated in Fig. \ref{mcufig}, which we will refer to as a multiport-based control unit (MCU). For the moment, no additional phase shifts are assumed  for simplicity. More details of correlated photon amplitude manipulation can be found in \cite{osawa2020higher}. In  Fig. \ref{fig:review_operation}, two photons entering this configuration from the left with no phase shifts are transformed according to:
\begin{eqnarray}
e_0f_0 &\xrightarrow{BS}& \frac{1}{2}(a_0-b_0)(a_0+b_0)\nonumber \\
&\xrightarrow{M}& \frac{1}{2}(a_0-b_0)(c_0+d_0) \xrightarrow{BS} e_0f_1.
\end{eqnarray}
The initial state $e_0f_0$ depicts a configuration in which two correlated photons are introduced from the left side, one into each mode of the first beam splitter.
The multiport (M) operation splits incoming photon amplitudes into right moving and left moving amplitudes distributed over two rails (waveguides). After that, the state distributed between two rails and transformed by the multiport amplitudes again enters beam splitters on both sides and photons reappear as a correlated two-photon state but now at different spatial locations (modes) - one on the right and one on the left as illustrated in Fig. \ref{fig:review_operation}. The workings of all components are assumed here to be identical for both horizontal and vertical polarizations. This feature makes it possible to generalize such linear-optical state transformation manipulation on the large family of polarization-entangled states and opens an exciting opportunity for constructing large-scale reconfigurable entanglement distribution networks.

This article is organized as follows. In Sec. \ref{sec:quantum_state_distribution}, we introduce the major principles governing the polarization-entangled quantum state manipulation in a multiport unit - the central element of the network.
We show that entangled two-photon states can be redistributed between multiple spatial modes.
It is shown in Sec. \ref{sec:scalable} how the use of passive phase shifters and sequential application of several multiport units enables covering a much greater number of output modes and sharing entanglement between desired end-users. 
The final goal of developing a fully reconfigurable optical quantum network could only be reached when the controllable change in direction of amplitude propagation is attainable.
Sec. \ref{sec:active} describes how the use of controllable time-dependent active phase shifters enables this important feature and opens the way to  building scalable multiple-user reconfigurable quantum entanglement distribution network structure. The summary is presented in Sec. \ref{sec:conclusion}.

\section{Controllable entangled-photon state distribution} \label{sec:quantum_state_distribution}

\begin{figure}[t]
	\centering
	\includegraphics[clip,width=\columnwidth]{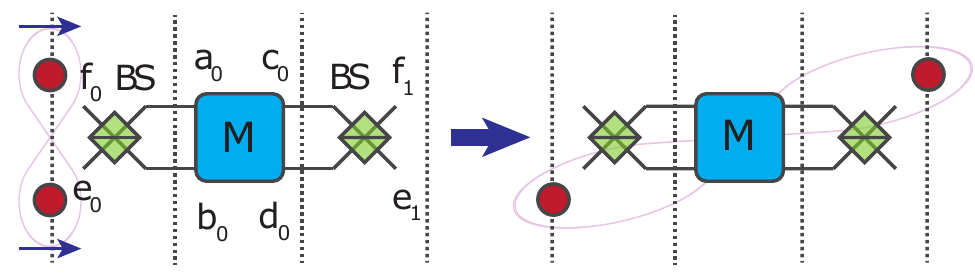}
	\caption{Entanglement distribution in a multiport unit.
	Two polarization-entangled photons occupying spatial modes $e_0$ and $f_0$ are inserted with the propagation direction to the right. The photons leave through different ports $e_0$ and $f_1$ after the linear-optical transformation at the multiport unit and sustain their polarization entanglement. The figure-eight shape is used here and throughout the rest of the paper to show entanglement between pairs of photons.
	}
	\label{fig:entanglement_distribution}
\end{figure}

The experimental setup illustrating the operation of the central unit in the proposed network, the MCU, on a polarization-entangled state is illustrated in Fig. \ref{fig:entanglement_distribution}. Throughout this paper, the figure-eight shape shown in this diagram will be used to denote entanglement between the enclosed pair of photons.
The MCU will be used as a building block to enable distribution of two-photon entanglement as well as multi-photon entangled states among any desired  spatial modes. In the case of an entangled-photon pair, photons enter at $e_0$ and $f_0$ and propagate towards the right (see Fig. \ref{fig:entanglement_distribution}, left).

\subsection{Manipulation of entangled two-photon states}

\begin{figure}[b]
\centering
\subfloat[]{\includegraphics[clip,width=\columnwidth]{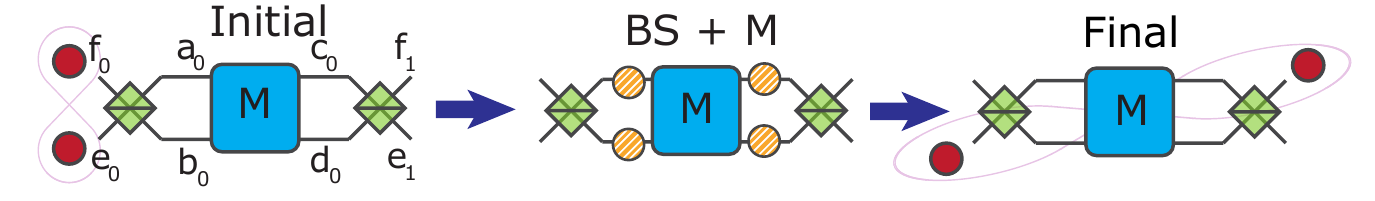}\label{}}\\
\vspace{-15px}
\subfloat[]{\includegraphics[clip,width=\columnwidth]{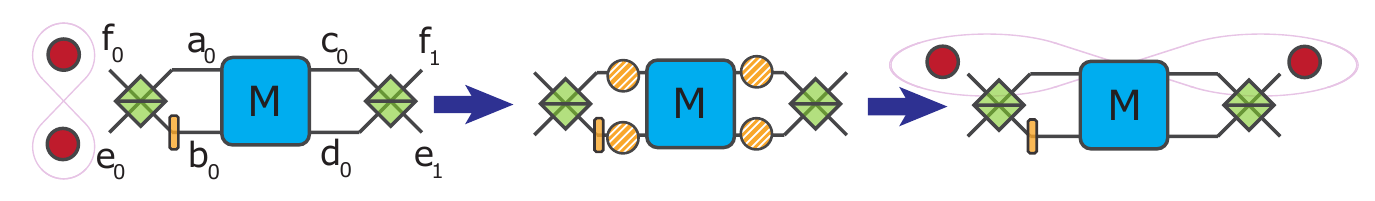}\label{}}\\
\vspace{-15px}
\subfloat[]{\includegraphics[clip,width=\columnwidth]{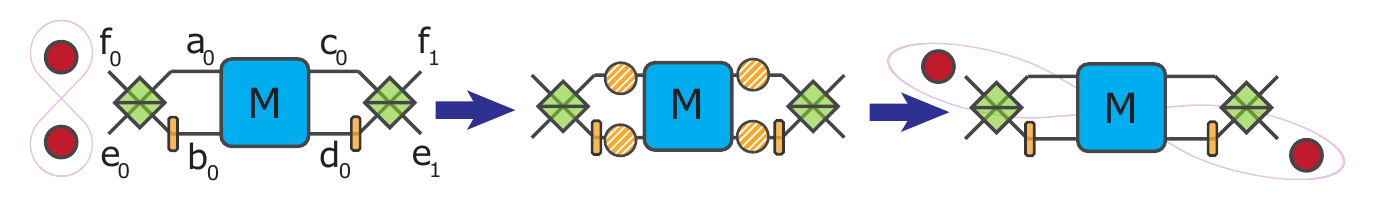}\label{}}\\
\vspace{-15px}
\subfloat[]{\includegraphics[clip,width=\columnwidth]{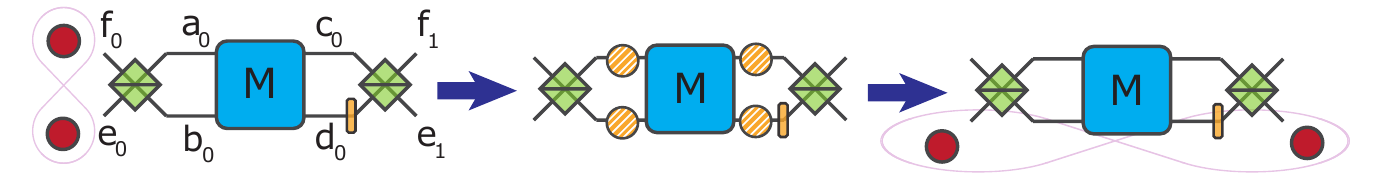}\label{}}
\caption{
Entangled photon exit amplitude control. The input photons enter from ports $e_0$ and $f_0$. Depending on the placement of the phase shifters, the photon pair could emerge at four different spatial modes.}
     \label{fig:two_photon_entangled}
\end{figure}

We analyze the distribution of two-photon entangled Bell states exploiting polarization degrees of freedom.
Initially, the input polarization-entangled state is transformed by a beam splitter; then the transformed input state is processed by the multiport. The multiport splits the input state into right-moving and left-moving amplitudes confined to a two-rail waveguided path (striped orange circles in Fig. \ref{fig:two_photon_entangled}).

We introduce polarization-entangled Bell states in both bra-ket and creation operator notation. We use the simplified notation for creation operators by removing hat and dagger symbols.
The creation operators are always acting on vacuum state $\ket{0}$, therefore, we often omit writing the vacuum states out explicitly.
\begin{eqnarray}
    \ket{\psi^\pm} &=& \frac{1}{\sqrt{2}}(\ket{H}_0\ket{V}_0\pm\ket{V}_0\ket{H}_0) \nonumber \\
    &=& \frac{1}{\sqrt{2}}(e_{H0}f_{V0} \pm e_{V0}f_{H0})\ket{0},\\
    \ket{\phi^\pm} &=& \frac{1}{\sqrt{2}}(\ket{H}_0\ket{H}_0\pm\ket{V}_0\ket{V}_0) \nonumber \\
    &=&\frac{1}{\sqrt{2}}(e_{H0}f_{H0} \pm e_{V0}f_{V0}) \ket{0}.
\end{eqnarray}

Initially, the Bell states are transformed by a beam splitter so that their amplitudes are spread between two spatial waveguides:
\begin{eqnarray}
\ket{\phi^{\pm}} &\xrightarrow{BS}& \frac{1}{2\sqrt{2}}({a_{H0}}^2 \pm {a_{V0}}^2-{b_{H0}}^2 \mp {b_{V0}}^2)\ket{0},\\
\ket{\psi^+} &\xrightarrow{BS}& \frac{1}{\sqrt{2}}(a_{H0}a_{V0}-b_{H0}b_{V0})\ket{0},\\
\ket{\psi^-} &\xrightarrow{BS}& \frac{1}{\sqrt{2}}(a_{H0}b_{V0}-a_{V0}b_{H0})\ket{0}.
\end{eqnarray}

Then two input state amplitudes occupying  spatial modes $a_0$ and $b_0$ enter the Grover multiport from the left and get distributed between the right side and left side of the multiport without any losses in the system.  The final amplitude transformation is performed by two beam-splitters on either side of the Grover multiport. A detailed description of the  amplitude transformation at the beam splitter and at the Grover multiport could be found in \cite{osawa2020higher}.
We provide detailed calculation on the transformation of $\ket{\phi^+}$.

\begin{eqnarray}
\ket{\phi^+} &\xrightarrow{BS}& \frac{1}{2\sqrt{2}}(a_{H0}^2+a_{V0}^2-b_{H0}^2-b_{V0}^2)\ket{0} \nonumber \\
&\xrightarrow{M}& -\frac{1}{2\sqrt{2}}((a_{H0}-b_{H0})(c_{H0}+d_{H0})\nonumber \\
&& \qquad +(a_{V0}-b_{V0})(c_{V0}+d_{V0}))\ket{0} \nonumber \\
&\xrightarrow{BS}& -\frac{1}{\sqrt{2}}(e_{H0}f_{H1}+e_{V0}f_{V1})\ket{0}.
\end{eqnarray}
The original input entanglement between $e_0$ and $f_0$ modes is now redistributed between two modes $e_0$ and $f_1$. These new modes become entangled as a result of the multiport-based control unit (MCU) action. Such a distribution feature is valid for all four polarization Bell states and can be demonstrated by engaging the same amplitude transformation procedure.
\begin{eqnarray}
\ket{\phi^-} &\xrightarrow{MCU}& \frac{1}{\sqrt{2}}(e_{H0}f_{H1}-e_{V0}f_{V1})\ket{0}, \nonumber \\
\ket{\psi^{\pm}} &\xrightarrow{MCU}& \frac{1}{\sqrt{2}}(e_{H0}f_{V1} \pm e_{V0}f_{H1})\ket{0}. \end{eqnarray}

\subsection{Control of exit modes}
\label{sec:exit_control}
The final spatial location of exit photons belonging to the same entangled pair after their manipulation by the linear-optical multiport unit could be controlled by imposing a $\pi$ phase shift denoted as $P$ on a particular communication waveguide \cite{osawa2020higher}.
This enables one to control the propagation of entangled state amplitudes and guides its energy to the desired network nodes.

Consider $\ket{\phi^+}$ Bell state again, with the state introduced in the multiport unit from the left at $e_{0}f_{0}$ ports.
The input output relationship is give in Fig. \ref{fig:two_photon_entangled} (a).
We show the propagation amplitude can be changed by introducing the $\pi$ phase shift (P) on the right side of the multiport unit indicated in Fig. \ref{fig:two_photon_entangled} (d).
\begin{eqnarray}
&\ket{\phi^+} = \frac{1}{\sqrt{2}}(e_{H0}f_{H0}+e_{V0}f_{V0})\ket{0} \nonumber \\
\xrightarrow{BS+M} &-\frac{1}{2\sqrt{2}}((a_{H0}-b_{H0})(c_{H0}+d_{H0})\nonumber \\
&+(a_{V0}-b_{V0})(c_{V0}+d_{V0}))\ket{0}.
\label{eqn:example_state}
\end{eqnarray}
This state is now a superposition illustrating coupled right-moving and left-moving amplitudes.
Each term is transformed once more by the final beam splitter interaction,
\begin{eqnarray}
(a_{0}-b_{0}) &\xrightarrow{BS}& e_0, \label{eqn:minus_ab}\\
(a_{0}+b_{0}) &\xrightarrow{BS}& f_0, \label{eqn:plus_ab}\\
(c_{0}-d_{0}) &\xrightarrow{BS}& e_1, \label{eqn:minus_cd}\\
(c_{0}+d_{0}) &\xrightarrow{BS}& f_1. \label{eqn:plus_cd}
\end{eqnarray}

We dropped scaling coefficients since they are irrelevant for this discussion of relative phase effects.

If the relative phase between the two rails on both sides of the multiport is positive, as in Eq. (\ref{eqn:plus_ab}) and Eq. (\ref{eqn:plus_cd}), then both photons leave the unit at ports $f_0$ and $f_1$.
Similarly, if the relative phase between the two rails on both sides of the multiport is negative, as in Eq. (\ref{eqn:minus_ab}) and Eq. (\ref{eqn:minus_cd}), then both photons leave at ports $e_0$ and $e_1$.
We can establish the following correspondence between the phase shifts between upper and lower rails on a particular part of the multiport (left phase, right phase) and exit ports for a particular portion of the entangled state:

\begin{eqnarray} (+,+)=(f_0,f_1), &\qquad \qquad & (+,-)=(f_0,e_1),\\  (-,+)=(e_0,f_1), &\qquad \qquad & (-,-)=(e_0,e_1).\end{eqnarray} The choice of phase shifts therefore provides complete control over the output distribution of the entangled state spatial modes.

We take the final line of Eq. (\ref{eqn:example_state}) and obtain a state $\ket{\xi}$ amplitude distribution prior to encountering final beam splitters,
\begin{eqnarray}
\ket{\xi} \equiv -\frac{1}{2\sqrt{2}}((a_{H0}-b_{H0})(c_{H0}+d_{H0})\nonumber \\
\qquad+(a_{V0}-b_{V0})(c_{V0}+d_{V0}))\ket{0}
\label{eqn:redefined_state}
\end{eqnarray}
Eq. (\ref{eqn:redefined_state}) has a form of (-,+) for the first term, and (-,+) for the second term.
The final BS transformation on both sides provides
\begin{eqnarray}
\ket{\xi} \xrightarrow{BS} \frac{1}{\sqrt{2}}(e_{H0}f_{H1}+e_{V0}f_{V1})\ket{0}.
\end{eqnarray}
A phase shift in the right side (see Fig. \ref{fig:two_photon_entangled} (d)) can alter Eq. (\ref{eqn:redefined_state}) from (-,+) to (-,-) for the first term and from (-,+) to (-,-) for the second term thus resulting in entangled amplitudes exiting from different ports of the apparatus.
The result after the phase alteration is
\begin{eqnarray}
\ket{\xi} \xrightarrow{P+BS} \frac{1}{\sqrt{2}}(e_{H0}e_{H1}+e_{V0}e_{V1})\ket{0}.
\end{eqnarray}
In other words, the exiting photon in spatial mode $f_{H1}$ is switched now to the mode $e_{H1}$ and $f_{V1}$ is switched with $e_{V1}$.
The result indicates we can actively redirect the amplitude propagation direction without disturbing the entanglement.

\section{Scalability of entangled-state distribution network} \label{sec:scalable}
\begin{figure}[t]
\centering
\subfloat[]{\includegraphics[clip,width=\columnwidth]{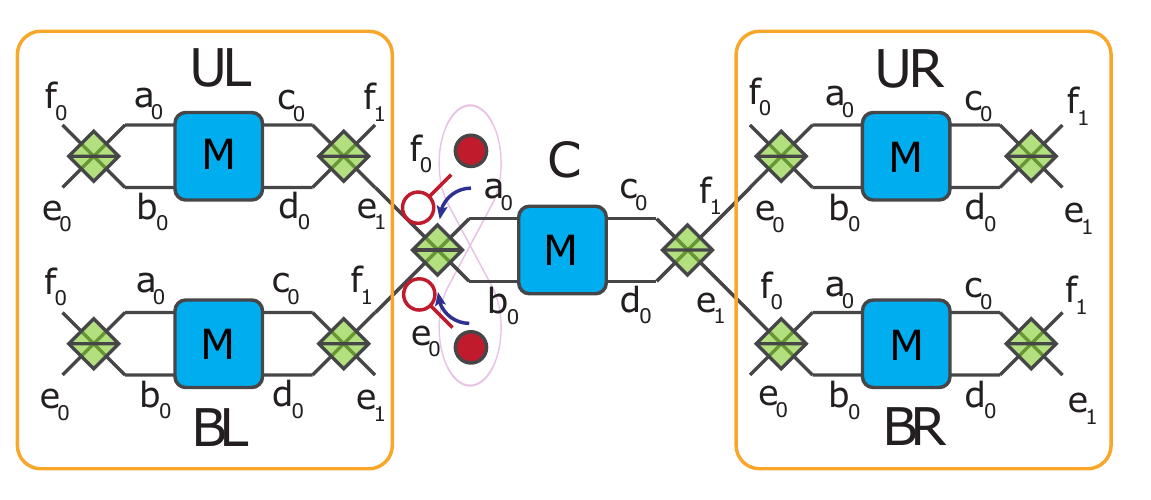}\label{}}\\
\subfloat[]{\includegraphics[clip,width=\columnwidth]{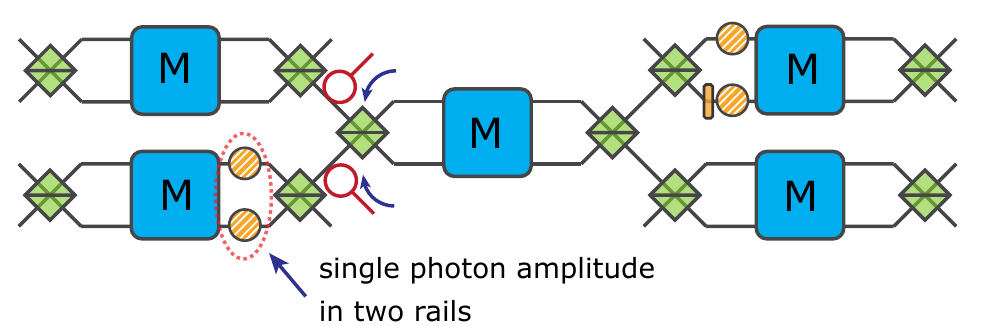}\label{}}\\
\subfloat[]{\includegraphics[clip,width=\columnwidth]{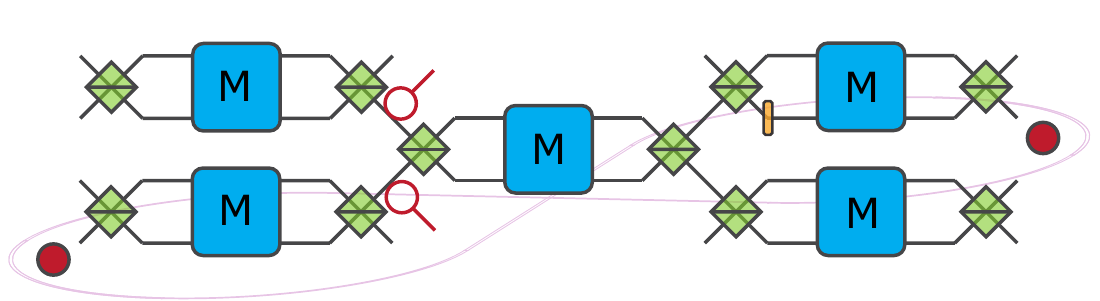}\label{}}
\caption{
Two-photon entanglement distribution in a network. (a)Network schematic and initial conditions. The network consists of five multiport-based units. Entangled photons are inserted at the central unit C. After the transformation in the first multiport unit, the photons are propagated farther to the second layer of the network.
(b)Amplitude propagation after the first round of multiport transformation. Second layer of the network. The photons leaving C enter UL and BR units in this case and get transformed by beam splitters.  The two-photon amplitudes (yellow striped circles) travel through the second layer of multiport devices.
(c) Final outcome for the two-photon entanglement propagation through the network structure. Amplitudes are transformed again by final beam splitters resulting in the distribution of the original entangled state energy between the right and left side of the system. }
     \label{fig:entanglement_network}
\end{figure}

To further increase the number of spatial modes, we can link several multiport-based control units together.
A network could be composed of several units of the type previously discussed, each encompassing two beam splitters and a multiport device, possibly with phase shifters.
Such units can be arranged in a network structure by connecting their ports using on-chip waveguides or by fiber links when it is desirable to cover longer distances.
A schematic setup in Fig. \ref{fig:entanglement_network} illustrates the two-photon quantum-entangled state transportation between 8 user nodes (16 modes) using multiport units.
For illustration purpose, this network is separated into five sections.
C, UL, BL, UR, and BR are central upper left, bottom left, upper right, and bottom right multiport unit locations, respectively.
A two-photon entangled state is introduced from the left side into the central network portion using optical isolators.
\begin{eqnarray}
e_{0C} f_{0C} &\xrightarrow{BS}&-\frac{1}{2}(a_{0C}^2-b_{0C}^2) \nonumber \\
&\xrightarrow{M}& -\frac{1}{2}(a_{0C}-b_{0C})(c_{0C}+d_{0C}) \nonumber \\
&\xrightarrow{BS}& e_{0C} f_{1C}.
\end{eqnarray}
This equation illustrates a two-photon state distribution between spatial modes at the end of the central part of the system action.
Such transformed photons are now entering the bottom left and the upper right second layer units because $e_{0C}$ is propagating to the left and $f_{1C}$ is propagating to the right.
\begin{eqnarray}
&\xrightarrow{BS}& \frac{1}{2}(c_{0UL}+d_{0UL})(a_{0BR}-b_{0BR}) \nonumber \\
&\xrightarrow{P+M}& \frac{1}{2}(a_{0UL}+b_{0UL})(c_{0BR}+d_{0BR})\nonumber \\ 
&\xrightarrow{BS}& e_{0UL}e_{1BR},
\end{eqnarray}


where $e_{0C}$ indicates a photon in the central multiport region occupying spatial mode $e_0$.
Subscripts UL, BL, UR, and BR are given for appropriate creation operators.

The photon propagation direction can be controlled by inserting phase shifters, as described in Sec. \ref{sec:exit_control}.
This capability holds for both photon polarizations.
The quantum state transformation pattern (photon $\to$ distributed two-rails amplitudes $\to$ photon) repeats every time a photon leaves the multiport unit and enters another one.
Fig. \ref{fig:entanglement_network} illustrates the transition of the two-photon state energy from the central multiport to the bottom left and top right multiports (Fig. \ref{fig:entanglement_network} (c)).
The form of the quantum state is preserved along its propagation, and the entanglement distribution between any of the two exit locations of the entire system can be performed.
We can choose any of the four outputs on the right side for the right moving amplitude and any of the four outputs on the left side for the left moving amplitude by utilizing fixed phase shifters.
\subsection{Entanglement distribution control using passive phase shifters}

\begin{figure}[bt]
	\centering
	\includegraphics[width=3.5in]{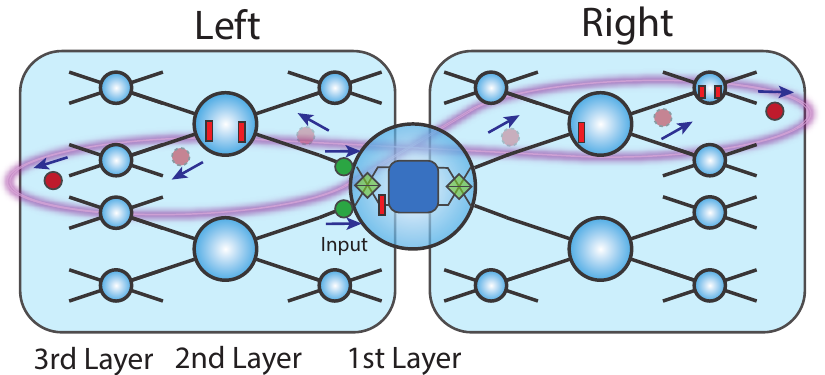}
	\caption{Amplitude control using passive phase shifters. Two photons are sent from the left side of the central multiport device (green circles). The red square in the circle indicates phase shifters embedded in the multiport unit. The phase shifters change the  propagation direction in transverse direction as described in Section IIB. The final outgoing photon locations are illustrated by solid red circles. The amplitude distribution during internal transport though the network is shown by faded red circles.
	}
	\label{fig:scaled_network_ex_1}
\end{figure}

The network scalability is illustrated in Fig. \ref{fig:scaled_network_ex_1} by a three-layered configuration of multiport units.
We can impose additional control to direct entangled photons towards specific outgoing spatial modes in the network by inserting fixed phase shifters at each multiport location.

One specific example of this network operation is provided in Fig. \ref{fig:scaled_network_ex_1}.
After leaving the first multiport unit, the photon enters the next multiport unit either from the top side mode or the bottom side mode.
When the photon enters from the bottom, the relative phase difference between the two output modes is -1 (a-b).
This means that the amplitude is going to be reflected at the second multiport back to the original location if no phase shifter is introduced in the path.
The introduction of $\pi$ phase shift allows the amplitude to keep propagating forward.
If the photon enters from the top mode of the beam splitter then, the relative phase remains +1 (a+b), therefore the photon propagates forward  without any additional phase shifts.

\begin{figure*}[tbh]
	\centering
\subfloat[]{\includegraphics[clip,width=1.75\columnwidth]{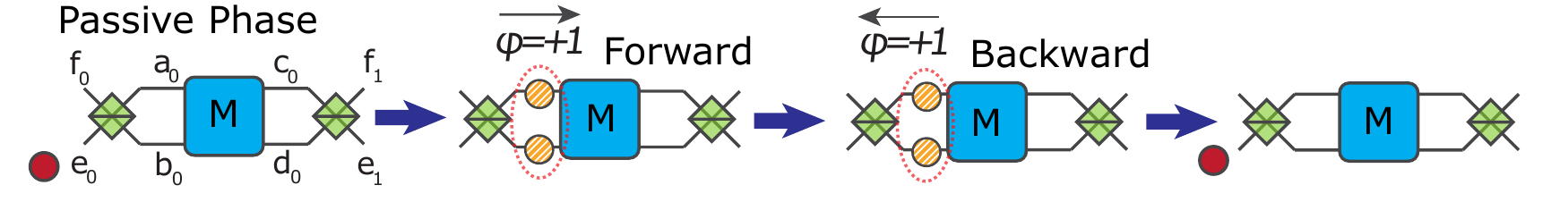}\label{}}\\
\vspace{-15px}
\subfloat[]{\includegraphics[clip,width=1.75\columnwidth]{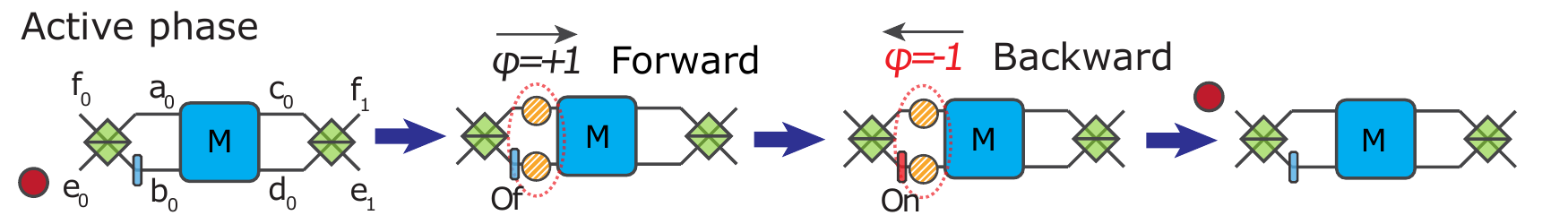}\label{}}\\
	\caption{Passive and active phase shift altering propagation direction of reflected amplitude. (a) Passive phase shifts. In this case, we do not need to apply any phases therefore there is no phase shifters in the unit. A single photon is entering from the mode $e_0$. The photon is transformed by the BS into a superposition of forward propagating amplitudes towards the multiport device. Depending on the relative phase shift between the two rails, the amplitude gets either transmitted or reflected at the multiport. The original amplitude transformation done by the input  BS for the input mode $e$ has negative relative phase, therefore it will reflect back from the multiport device if one apply zero phase shift (denoted as $\phi=+1$). The second passage through the passive $\phi=+1$ passive phase shifter dictates the photon will exit at the same entrance mode $e_0$.
	(b) The active switch of the phase shifter to $\phi=-1$ only for the backward propagating amplitudes changes the relative phase between two rails and switches the outgoing photon energy to the $f_0$ mode. The active phase shifter allows switching spatial propagation modes $e_0 \rightarrow f_0$ while  the passive phase shift only allows a full return $e_0 \rightarrow e_0$. Additional $\pi$ phase is added to all phase shifters if the photon is entering from the port $f_0$ and is getting switched to $e_0$.
	}
	\label{fig:active_e}
\end{figure*}

Let us consider the amplitude evolution on the right side.
The detailed discussion of the amplitude propagation control using phase shifts is provided earlier in Sec. \ref{sec:exit_control}.
The photon departs the first unit from the bottom side and enters the next unit from the top side of the BS.
The transformed amplitude has a negative relative phase, hence the phase must be altered to make sure the photon propagates forward.
Then, the propagated amplitude can leave the second layer either from the upper or lower side of the unit depending on the relative phase settings.

Assume that the amplitude leaves the second layer at the upper mode since the photon has a positive relative phase.
The same procedure is repeated at the third layer of multiport units except for the final phase shift.
The final phase shift in the third stage is set to guide the photon energy to emerge from the lower outgoing port of the unit.
The same procedure applies to the left side amplitude as well.

\section{Reconfigurable networks and controllable entanglement distribution} \label{sec:active}

\subsection{Redirection of amplitude propagation using active phase shifters} \label{sec:active_phase_A}

We have described state manipulation methods by using only passive phase shifters up to this point.
The passive phase shifter scheme is limited in performance and does not allow for an arbitrary redirection of photons in the multi-user network.
Each photon from an entangled pair may independently experience multiport unit transformations while traveling in a complex network.
Consider the passive phase shifter based network again (see. Fig. \ref{fig:scaled_network_ex_1}) and focus on the left side of the network.
In the passive network (constant phase shifts), the second layer branches out into four outputs, two in forward direction and two for a backward direction.

When a passive network is used, it is impossible for the input photon to exit from the mode $f_0$ in order to modify the general propagation path in the network.  This makes 20 out of 36 nodes in such passive network unattainable.
One must engage active phase shifters (phase shifts that can be controllably changed during the operation of the network) in order to reverse the photon propagation direction from mode $e_0$ to $f_0$ as well as from mode $f_0$ to $e_0$.
A single-photon transformation using active phase shifters is presented  that enables a fully reconfigurable network infrastructure.

We focus on the propagation of one part of the two-photon entangled state traveling over the network and its active amplitude manipulation (see Fig.  \ref{fig:active_e}).
A single photon entering port $e_0$ can leave either at $e_0$ or $f_0$ depending on the phase manipulation.
We compare the amplitude manipulation using both passive shifts and active shifts.

The system introduces zero forward passive phase shift and zero backward phase shift as indicated in Fig. \ref{fig:active_e} (a).
The photon is sent from mode e, therefore there is no need for a phase shift to experience reflection at the multiport device in this case.

This can be seen explicitly by looking at the transformation of the state operators. The beam splitter transformation on a single photon from mode $e_0$ is,
\begin{eqnarray}
e_0 \xrightarrow{BS} \frac{1}{\sqrt{2}}(a_0-b_0).
\end{eqnarray}
The transformed amplitude experiences zero phase shift to keep the relative phase equal to $-1$.
\begin{eqnarray}
\xrightarrow{P}\frac{1}{\sqrt{2}}(a_0-b_0).
\end{eqnarray}
The amplitude gets reflected at the multiport device,
\begin{eqnarray}
\xrightarrow{M} \frac{1}{\sqrt{2}}(a_0-b_0).
\end{eqnarray}
A passive phase shifter cannot change the amount of phase in a time dependent manner, therefore the backward propagation experiences the same amount of phase (zero in this case).
Finally, the reflected amplitude is transformed again by the BS resulting in the same location as the initial input location.
\begin{eqnarray}
\xrightarrow{P}\frac{1}{\sqrt{2}}(a_0-b_0)\xrightarrow{BS} e_0.
\end{eqnarray}

Consider the action of an active phase shifter as provided in Fig. \ref{fig:active_e} (b).
Now the system introduces zero phase shift for a forward amplitude and $\pi$ phase shift after the multiport reflection thus allowing to switch the photon propagation from $e_0$ to $f_0$.
The first three operations are the same as the passive phase shifter case,
\begin{eqnarray}
e_0 \xrightarrow{BS+P+M} \frac{1}{\sqrt{2}}(a_0-b_0).
\end{eqnarray}
After the multiport transformation, the exit mode will be different from the input mode if $\pi$ phase shift is introduced.
The time-dependent active phase manipulation is required to execute different phase shifts.
\begin{eqnarray}
\xrightarrow{P} \frac{1}{\sqrt{2}}(a_0+b_0) \xrightarrow{BS} f_0.
\end{eqnarray}
The amplitude is redirected to mode $f_0$ using active phase shifters.

\subsection{Centralized network}

When passive phase shifters are used, the reflected amplitudes simply propagate back to the original input locations.  Active phase shifters enable full propagation control of reflected amplitudes as well as transmitted ones.

\begin{figure}[tbh]
	\centering
\includegraphics[width=3.5in]{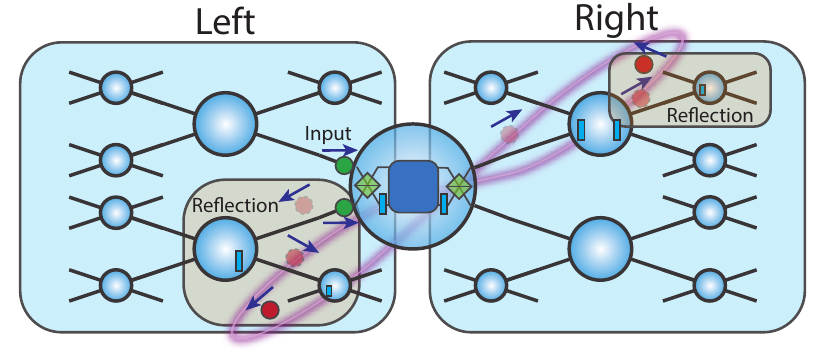}
\caption{Full amplitude distribution control using active phase shifters. Two photons (green circles) are inserted from the left side of the multiport device. The blue square in the circle indicates active phase shifters embedded in the unit. The time-dependent phase shifters change the propagation direction when it is required.
The final photon location is illustrated as solid red circles. The path towards the final destination is traced by faded red circles.
The reflection to any desired port is now available in an active network.
The reflection occurs in  sections indicated by gray boxes.}
\label{fig:scaled_network_ex_2}
\end{figure}

Consider an example of the entangled state distribution illustrated in Fig. \ref{fig:scaled_network_ex_2}.
The network configuration is similar to the one in Fig. \ref{fig:scaled_network_ex_1} but all passive phase shifters are replaced by active phase shifters now.
The initial state is split into right moving and left moving amplitudes by the central multiport unit.
The left moving amplitude leaving port $f_0$ while the right moving amplitude leaves port $e_1$ of the central unit.
The amplitude leaving $f_0$ enters one of the second layer unit from the port $e$.
The active phase shifter is now required in the second layer to switch the photon direction from port $e$ to $f$ on the same side after its reflection at the multiport unit. 
The photon proceed to the third layer after an appropriate phase shift at the second layer.
Unlike the passive network, all inner third layer ports become accessible now.
The photon entering the third layer from port $e$ leaves the system after reflection again.
The right moving amplitude can be controlled in a similar way as the left moving amplitude.
The right moving amplitude leaves from port $e$ of the central unit.
The amplitude enters a unit in the second layer from the port $f$.
Unlike the left moving amplitude, the relative phase between two rails is positive since the amplitude is entered from $f$.
The photon amplitude passes though the second layer multiport unit and leaves it from the $e$ port. One more reflection is introduced in the third layer by applying an active phase shift as described above.

With the use of active phase shifters embedded in the system, the number of output ports increases to 36 from 16 in a three-layer setup.
The scalability can be generalized to $4\times 3^n$ where n is the number of layers.
In this subsection, we only considered the centralized configuration of the network where independent amplitudes travel over the right side and the left side of the network. In general, the right side amplitude can be redirected back to the left side of the device so that the two amplitude can reach end nodes positioned on the same side of the system.

\subsection{Decentralized fully reconfigurable entanglement distribution network}

\begin{figure}[tbh]
\centering
\includegraphics[clip,width=\columnwidth]{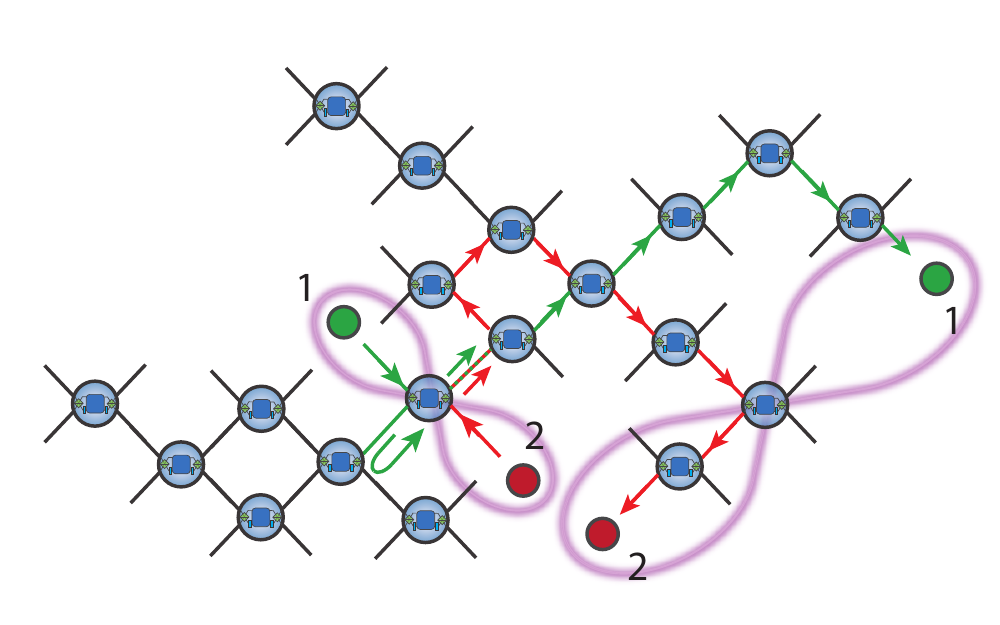}
\caption{Decentralized fully reconfigurable entanglement distribution network. Two entangled photons (red and green circles) are inserted from two ports of the multiport unit.
The propagation path of the first photon (labeled as 1) is indicated in green color and the second photon (labeled as 2) is indicated in red.
They are following colored arrows in the diagram and finally entangle quantum devices at two remote nodes.
}
\label{fig:decentralized_network_both}
\end{figure}

A completely flexible and reconfigurable network design can be implemented using the technical approach introduced above.
One can insert two entangled photons at any part of the network indicated in Fig. \ref{fig:decentralized_network_both}.
In this setup, we assume again that active phase shifters are available. There is no need for circulators at the insertion point now due to the active phase shifter availability.

The first photon of an entangled pair (labeled as 1 and displayed as a green circle) follows the green path, the second photon (labeled as 2, with a red circle) follows the red path.
The first photon gets reflected at the first multiport unit, and the photon is sent to a unit in the opposite direction of the second photon.
We set the active phase shifter so that the first photon is then reflected at the second multiport unit making the green photon to come back to the first unit.
The rest of the propagation follows the same rule as the centralized network.
Likewise, the second red photon gets reflected at the first multiport device and travels to the destination following the red arrows.
The second photon experiences reflection with change in propagation direction by active phase shifters three times.
The input state is traveling to two distinct locations and the output state remains entangled.
We assumed the system is lossless, and the entanglement transportation is only done by phase shifts. 
The scheme does not require any post-selection, hence there is no reduction of amplitudes.
We chose this specific travel path as an example.  
However the setup allows the users to engage any desired propagation pattern by introducing corresponding active phase mapping and redirecting  entangled photons to different remote locations.

Note that the u-turn between the second and third time steps was added for the green photon in Fig. \ref{fig:decentralized_network_both} in order to ensure that the two photons were not on the same edge at the same time. This allowed them to be controlled independently: by making sure they traverse any common links in their trajectories at different times, the two photons can be made to encounter different phase shifts at those locations and thereby be directed in different directions.

\vspace{10pt}
\section{Conclusion}
\label{sec:conclusion}

We have introduced routing schemes for entangled photons using only linear optical devices through phase shifts for exit spatial mode control.
The multiport-based control unit (MCU) consisting of BS, multiport and phase shifters is shown to allow the sharing of entangled photons between two desired locations in a controllable manner, with only phase shifts required to achieve full control.
The system capacity is scaled by connecting the simple units to form a network.
Within the network, we considered the actions of passive and active phase shifters for routing of states.
Active phase shifters allow engaging both forward and backward propagation with full amplitude controllability in a reconfigurable network.

Using network structures, it is possible to introduce high photon number entangled states since the simultaneous appearence of multiphotons at a multiport device does not disable the state distribution.
This multiphoton entangled state distribution could be applied to quantum secret sharing \cite{hillery1999quantum,tittel2001experimental,bell2014experimental,williams2019quantum}, as well as quantum state stabilization schemes based on decoherence-free subspaces \cite{lidar1998decoherence,walton2003decoherence}.

Any stray photons that are not identical to the original entangled pair entering the system at any input have an arbitrary amplitude distribution over two rails after encountering the Grover coin. 
The amplitudes eventually end up at some random place in the network.
This random amplitude can be considered as some ambient background noise.
Coincidence measurements eliminate such noise, therefore the final detection of routed entanglement is still prominent.
In order for identical stray photons to introduce errors, those photons need to enter the same input port as the entangled photons at the same time having the same frequency and polarization.


A quantum repeater approach using the network with assistance from a third party can be introduced if a significant increment of distance and purification of an entangled state is required. Finally, if the phase shifts are allowed to be frequency dependent, multiplexing of multiple polarization-entangled photon pairs directed along different paths through the network simultaneously becomes a possibility.


This research was supported by the National Science Foundation EFRI-ACQUIRE Grant No. ECCS-1640968, AFOSR Grant No.FA9550-18-1-0056, and DOD/ARL Grant No.W911NF2020127.

\bibliographystyle{apsrev4-2}
\bibliography{state_distribution}

\providecommand{\noopsort}[1]{}\providecommand{\singleletter}[1]{#1}%
\begin{thebibliography}{42}%
\makeatletter
\providecommand \@ifxundefined [1]{%
 \@ifx{#1\undefined}
}%
\providecommand \@ifnum [1]{%
 \ifnum #1\expandafter \@firstoftwo
 \else \expandafter \@secondoftwo
 \fi
}%
\providecommand \@ifx [1]{%
 \ifx #1\expandafter \@firstoftwo
 \else \expandafter \@secondoftwo
 \fi
}%
\providecommand \natexlab [1]{#1}%
\providecommand \enquote  [1]{``#1''}%
\providecommand \bibnamefont  [1]{#1}%
\providecommand \bibfnamefont [1]{#1}%
\providecommand \citenamefont [1]{#1}%
\providecommand \href@noop [0]{\@secondoftwo}%
\providecommand \href [0]{\begingroup \@sanitize@url \@href}%
\providecommand \@href[1]{\@@startlink{#1}\@@href}%
\providecommand \@@href[1]{\endgroup#1\@@endlink}%
\providecommand \@sanitize@url [0]{\catcode `\\12\catcode `\$12\catcode
  `\&12\catcode `\#12\catcode `\^12\catcode `\_12\catcode `\%12\relax}%
\providecommand \@@startlink[1]{}%
\providecommand \@@endlink[0]{}%
\providecommand \url  [0]{\begingroup\@sanitize@url \@url }%
\providecommand \@url [1]{\endgroup\@href {#1}{\urlprefix }}%
\providecommand \urlprefix  [0]{URL }%
\providecommand \Eprint [0]{\href }%
\providecommand \doibase [0]{https://doi.org/}%
\providecommand \selectlanguage [0]{\@gobble}%
\providecommand \bibinfo  [0]{\@secondoftwo}%
\providecommand \bibfield  [0]{\@secondoftwo}%
\providecommand \translation [1]{[#1]}%
\providecommand \BibitemOpen [0]{}%
\providecommand \bibitemStop [0]{}%
\providecommand \bibitemNoStop [0]{.\EOS\space}%
\providecommand \EOS [0]{\spacefactor3000\relax}%
\providecommand \BibitemShut  [1]{\csname bibitem#1\endcsname}%
\let\auto@bib@innerbib\@empty
\bibitem [{\citenamefont {Kimble}(2008)}]{kimble2008quantum}%
  \BibitemOpen
  \bibfield  {author} {\bibinfo {author} {\bibfnamefont {H.~J.}\ \bibnamefont
  {Kimble}},\ }\href@noop {} {\bibfield  {journal} {\bibinfo  {journal}
  {Nature}\ }\textbf {\bibinfo {volume} {453}},\ \bibinfo {pages} {1023}
  (\bibinfo {year} {2008})}\BibitemShut {NoStop}%
\bibitem [{\citenamefont {Pant}\ \emph {et~al.}(2019)\citenamefont {Pant},
  \citenamefont {Krovi}, \citenamefont {Towsley}, \citenamefont {Tassiulas},
  \citenamefont {Jiang}, \citenamefont {Basu}, \citenamefont {Englund},\ and\
  \citenamefont {Guha}}]{pant2019routing}%
  \BibitemOpen
  \bibfield  {author} {\bibinfo {author} {\bibfnamefont {M.}~\bibnamefont
  {Pant}}, \bibinfo {author} {\bibfnamefont {H.}~\bibnamefont {Krovi}},
  \bibinfo {author} {\bibfnamefont {D.}~\bibnamefont {Towsley}}, \bibinfo
  {author} {\bibfnamefont {L.}~\bibnamefont {Tassiulas}}, \bibinfo {author}
  {\bibfnamefont {L.}~\bibnamefont {Jiang}}, \bibinfo {author} {\bibfnamefont
  {P.}~\bibnamefont {Basu}}, \bibinfo {author} {\bibfnamefont {D.}~\bibnamefont
  {Englund}},\ and\ \bibinfo {author} {\bibfnamefont {S.}~\bibnamefont
  {Guha}},\ }\href@noop {} {\bibfield  {journal} {\bibinfo  {journal} {npj
  Quantum Inf.}\ }\textbf {\bibinfo {volume} {5}},\ \bibinfo {pages} {1}
  (\bibinfo {year} {2019})}\BibitemShut {NoStop}%
\bibitem [{\citenamefont {Duan}\ \emph {et~al.}(2001)\citenamefont {Duan},
  \citenamefont {Lukin}, \citenamefont {Cirac},\ and\ \citenamefont
  {Zoller}}]{duan2001long}%
  \BibitemOpen
  \bibfield  {author} {\bibinfo {author} {\bibfnamefont {L.-M.}\ \bibnamefont
  {Duan}}, \bibinfo {author} {\bibfnamefont {M.~D.}\ \bibnamefont {Lukin}},
  \bibinfo {author} {\bibfnamefont {J.~I.}\ \bibnamefont {Cirac}},\ and\
  \bibinfo {author} {\bibfnamefont {P.}~\bibnamefont {Zoller}},\ }\href@noop {}
  {\bibfield  {journal} {\bibinfo  {journal} {Nature}\ }\textbf {\bibinfo
  {volume} {414}},\ \bibinfo {pages} {413} (\bibinfo {year}
  {2001})}\BibitemShut {NoStop}%
\bibitem [{\citenamefont {Duan}\ and\ \citenamefont
  {Kimble}(2004)}]{duan2004scalable}%
  \BibitemOpen
  \bibfield  {author} {\bibinfo {author} {\bibfnamefont {L.-M.}\ \bibnamefont
  {Duan}}\ and\ \bibinfo {author} {\bibfnamefont {H.}~\bibnamefont {Kimble}},\
  }\href@noop {} {\bibfield  {journal} {\bibinfo  {journal} {Phys. Rev. Lett.}\
  }\textbf {\bibinfo {volume} {92}},\ \bibinfo {pages} {127902} (\bibinfo
  {year} {2004})}\BibitemShut {NoStop}%
\bibitem [{\citenamefont {Zhao}\ \emph {et~al.}(2009)\citenamefont {Zhao},
  \citenamefont {Dudin}, \citenamefont {Jenkins}, \citenamefont {Campbell},
  \citenamefont {Matsukevich}, \citenamefont {Kennedy},\ and\ \citenamefont
  {Kuzmich}}]{zhao2009long}%
  \BibitemOpen
  \bibfield  {author} {\bibinfo {author} {\bibfnamefont {R.}~\bibnamefont
  {Zhao}}, \bibinfo {author} {\bibfnamefont {Y.}~\bibnamefont {Dudin}},
  \bibinfo {author} {\bibfnamefont {S.}~\bibnamefont {Jenkins}}, \bibinfo
  {author} {\bibfnamefont {C.}~\bibnamefont {Campbell}}, \bibinfo {author}
  {\bibfnamefont {D.}~\bibnamefont {Matsukevich}}, \bibinfo {author}
  {\bibfnamefont {T.}~\bibnamefont {Kennedy}},\ and\ \bibinfo {author}
  {\bibfnamefont {A.}~\bibnamefont {Kuzmich}},\ }\href@noop {} {\bibfield
  {journal} {\bibinfo  {journal} {Nat. Phys.}\ }\textbf {\bibinfo {volume}
  {5}},\ \bibinfo {pages} {100} (\bibinfo {year} {2009})}\BibitemShut {NoStop}%
\bibitem [{\citenamefont {Nicolas}\ \emph {et~al.}(2014)\citenamefont
  {Nicolas}, \citenamefont {Veissier}, \citenamefont {Giner}, \citenamefont
  {Giacobino}, \citenamefont {Maxein},\ and\ \citenamefont
  {Laurat}}]{nicolas2014quantum}%
  \BibitemOpen
  \bibfield  {author} {\bibinfo {author} {\bibfnamefont {A.}~\bibnamefont
  {Nicolas}}, \bibinfo {author} {\bibfnamefont {L.}~\bibnamefont {Veissier}},
  \bibinfo {author} {\bibfnamefont {L.}~\bibnamefont {Giner}}, \bibinfo
  {author} {\bibfnamefont {E.}~\bibnamefont {Giacobino}}, \bibinfo {author}
  {\bibfnamefont {D.}~\bibnamefont {Maxein}},\ and\ \bibinfo {author}
  {\bibfnamefont {J.}~\bibnamefont {Laurat}},\ }\href@noop {} {\bibfield
  {journal} {\bibinfo  {journal} {Nat. Photon.}\ }\textbf {\bibinfo {volume}
  {8}},\ \bibinfo {pages} {234} (\bibinfo {year} {2014})}\BibitemShut {NoStop}%
\bibitem [{\citenamefont {Stephenson}\ \emph {et~al.}(2020)\citenamefont
  {Stephenson}, \citenamefont {Nadlinger}, \citenamefont {Nichol},
  \citenamefont {An}, \citenamefont {Drmota}, \citenamefont {Ballance},
  \citenamefont {Thirumalai}, \citenamefont {Goodwin}, \citenamefont {Lucas},\
  and\ \citenamefont {Ballance}}]{stephenson2020high}%
  \BibitemOpen
  \bibfield  {author} {\bibinfo {author} {\bibfnamefont {L.}~\bibnamefont
  {Stephenson}}, \bibinfo {author} {\bibfnamefont {D.}~\bibnamefont
  {Nadlinger}}, \bibinfo {author} {\bibfnamefont {B.}~\bibnamefont {Nichol}},
  \bibinfo {author} {\bibfnamefont {S.}~\bibnamefont {An}}, \bibinfo {author}
  {\bibfnamefont {P.}~\bibnamefont {Drmota}}, \bibinfo {author} {\bibfnamefont
  {T.}~\bibnamefont {Ballance}}, \bibinfo {author} {\bibfnamefont
  {K.}~\bibnamefont {Thirumalai}}, \bibinfo {author} {\bibfnamefont
  {J.}~\bibnamefont {Goodwin}}, \bibinfo {author} {\bibfnamefont
  {D.}~\bibnamefont {Lucas}},\ and\ \bibinfo {author} {\bibfnamefont
  {C.}~\bibnamefont {Ballance}},\ }\href@noop {} {\bibfield  {journal}
  {\bibinfo  {journal} {Phys. Rev. Lett.}\ }\textbf {\bibinfo {volume} {124}},\
  \bibinfo {pages} {110501} (\bibinfo {year} {2020})}\BibitemShut {NoStop}%
\bibitem [{\citenamefont {Englund}\ \emph {et~al.}(2010)\citenamefont
  {Englund}, \citenamefont {Shields}, \citenamefont {Rivoire}, \citenamefont
  {Hatami}, \citenamefont {Vuckovic}, \citenamefont {Park},\ and\ \citenamefont
  {Lukin}}]{englund2010deterministic}%
  \BibitemOpen
  \bibfield  {author} {\bibinfo {author} {\bibfnamefont {D.}~\bibnamefont
  {Englund}}, \bibinfo {author} {\bibfnamefont {B.}~\bibnamefont {Shields}},
  \bibinfo {author} {\bibfnamefont {K.}~\bibnamefont {Rivoire}}, \bibinfo
  {author} {\bibfnamefont {F.}~\bibnamefont {Hatami}}, \bibinfo {author}
  {\bibfnamefont {J.}~\bibnamefont {Vuckovic}}, \bibinfo {author}
  {\bibfnamefont {H.}~\bibnamefont {Park}},\ and\ \bibinfo {author}
  {\bibfnamefont {M.~D.}\ \bibnamefont {Lukin}},\ }\href@noop {} {\bibfield
  {journal} {\bibinfo  {journal} {Nano Lett.}\ }\textbf {\bibinfo {volume}
  {10}},\ \bibinfo {pages} {3922} (\bibinfo {year} {2010})}\BibitemShut
  {NoStop}%
\bibitem [{\citenamefont {Humphreys}\ \emph {et~al.}(2018)\citenamefont
  {Humphreys}, \citenamefont {Kalb}, \citenamefont {Morits}, \citenamefont
  {Schouten}, \citenamefont {Vermeulen}, \citenamefont {Twitchen},
  \citenamefont {Markham},\ and\ \citenamefont
  {Hanson}}]{humphreys2018deterministic}%
  \BibitemOpen
  \bibfield  {author} {\bibinfo {author} {\bibfnamefont {P.~C.}\ \bibnamefont
  {Humphreys}}, \bibinfo {author} {\bibfnamefont {N.}~\bibnamefont {Kalb}},
  \bibinfo {author} {\bibfnamefont {J.~P.}\ \bibnamefont {Morits}}, \bibinfo
  {author} {\bibfnamefont {R.~N.}\ \bibnamefont {Schouten}}, \bibinfo {author}
  {\bibfnamefont {R.~F.}\ \bibnamefont {Vermeulen}}, \bibinfo {author}
  {\bibfnamefont {D.~J.}\ \bibnamefont {Twitchen}}, \bibinfo {author}
  {\bibfnamefont {M.}~\bibnamefont {Markham}},\ and\ \bibinfo {author}
  {\bibfnamefont {R.}~\bibnamefont {Hanson}},\ }\href@noop {} {\bibfield
  {journal} {\bibinfo  {journal} {Nature}\ }\textbf {\bibinfo {volume} {558}},\
  \bibinfo {pages} {268} (\bibinfo {year} {2018})}\BibitemShut {NoStop}%
\bibitem [{\citenamefont {Kalb}\ \emph {et~al.}(2017)\citenamefont {Kalb},
  \citenamefont {Reiserer}, \citenamefont {Humphreys}, \citenamefont
  {Bakermans}, \citenamefont {Kamerling}, \citenamefont {Nickerson},
  \citenamefont {Benjamin}, \citenamefont {Twitchen}, \citenamefont {Markham},\
  and\ \citenamefont {Hanson}}]{kalb2017entanglement}%
  \BibitemOpen
  \bibfield  {author} {\bibinfo {author} {\bibfnamefont {N.}~\bibnamefont
  {Kalb}}, \bibinfo {author} {\bibfnamefont {A.~A.}\ \bibnamefont {Reiserer}},
  \bibinfo {author} {\bibfnamefont {P.~C.}\ \bibnamefont {Humphreys}}, \bibinfo
  {author} {\bibfnamefont {J.~J.}\ \bibnamefont {Bakermans}}, \bibinfo {author}
  {\bibfnamefont {S.~J.}\ \bibnamefont {Kamerling}}, \bibinfo {author}
  {\bibfnamefont {N.~H.}\ \bibnamefont {Nickerson}}, \bibinfo {author}
  {\bibfnamefont {S.~C.}\ \bibnamefont {Benjamin}}, \bibinfo {author}
  {\bibfnamefont {D.~J.}\ \bibnamefont {Twitchen}}, \bibinfo {author}
  {\bibfnamefont {M.}~\bibnamefont {Markham}},\ and\ \bibinfo {author}
  {\bibfnamefont {R.}~\bibnamefont {Hanson}},\ }\href@noop {} {\bibfield
  {journal} {\bibinfo  {journal} {Science}\ }\textbf {\bibinfo {volume}
  {356}},\ \bibinfo {pages} {928} (\bibinfo {year} {2017})}\BibitemShut
  {NoStop}%
\bibitem [{\citenamefont {Christandl}\ \emph {et~al.}(2004)\citenamefont
  {Christandl}, \citenamefont {Datta}, \citenamefont {Ekert},\ and\
  \citenamefont {Landahl}}]{christandl2004perfect}%
  \BibitemOpen
  \bibfield  {author} {\bibinfo {author} {\bibfnamefont {M.}~\bibnamefont
  {Christandl}}, \bibinfo {author} {\bibfnamefont {N.}~\bibnamefont {Datta}},
  \bibinfo {author} {\bibfnamefont {A.}~\bibnamefont {Ekert}},\ and\ \bibinfo
  {author} {\bibfnamefont {A.~J.}\ \bibnamefont {Landahl}},\ }\href@noop {}
  {\bibfield  {journal} {\bibinfo  {journal} {Phys. Rev. Lett.}\ }\textbf
  {\bibinfo {volume} {92}},\ \bibinfo {pages} {187902} (\bibinfo {year}
  {2004})}\BibitemShut {NoStop}%
\bibitem [{\citenamefont {Chapman}\ \emph {et~al.}(2016)\citenamefont
  {Chapman}, \citenamefont {Santandrea}, \citenamefont {Huang}, \citenamefont
  {Corrielli}, \citenamefont {Crespi}, \citenamefont {Yung}, \citenamefont
  {Osellame},\ and\ \citenamefont {Peruzzo}}]{chapman2016experimental}%
  \BibitemOpen
  \bibfield  {author} {\bibinfo {author} {\bibfnamefont {R.~J.}\ \bibnamefont
  {Chapman}}, \bibinfo {author} {\bibfnamefont {M.}~\bibnamefont {Santandrea}},
  \bibinfo {author} {\bibfnamefont {Z.}~\bibnamefont {Huang}}, \bibinfo
  {author} {\bibfnamefont {G.}~\bibnamefont {Corrielli}}, \bibinfo {author}
  {\bibfnamefont {A.}~\bibnamefont {Crespi}}, \bibinfo {author} {\bibfnamefont
  {M.-H.}\ \bibnamefont {Yung}}, \bibinfo {author} {\bibfnamefont
  {R.}~\bibnamefont {Osellame}},\ and\ \bibinfo {author} {\bibfnamefont
  {A.}~\bibnamefont {Peruzzo}},\ }\href@noop {} {\bibfield  {journal} {\bibinfo
   {journal} {Nat. Commun.}\ }\textbf {\bibinfo {volume} {7}},\ \bibinfo
  {pages} {1} (\bibinfo {year} {2016})}\BibitemShut {NoStop}%
\bibitem [{\citenamefont {Xu}\ \emph {et~al.}(2020)\citenamefont {Xu},
  \citenamefont {Zhang}, \citenamefont {Kong}, \citenamefont {Wang},\ and\
  \citenamefont {Long}}]{xu2020frequency}%
  \BibitemOpen
  \bibfield  {author} {\bibinfo {author} {\bibfnamefont {X.-S.}\ \bibnamefont
  {Xu}}, \bibinfo {author} {\bibfnamefont {H.}~\bibnamefont {Zhang}}, \bibinfo
  {author} {\bibfnamefont {X.-Y.}\ \bibnamefont {Kong}}, \bibinfo {author}
  {\bibfnamefont {M.}~\bibnamefont {Wang}},\ and\ \bibinfo {author}
  {\bibfnamefont {G.-L.}\ \bibnamefont {Long}},\ }\href@noop {} {\bibfield
  {journal} {\bibinfo  {journal} {Photonics Res.}\ }\textbf {\bibinfo {volume}
  {8}},\ \bibinfo {pages} {490} (\bibinfo {year} {2020})}\BibitemShut {NoStop}%
\bibitem [{\citenamefont {Perez-Leija}\ \emph {et~al.}(2013)\citenamefont
  {Perez-Leija}, \citenamefont {Keil}, \citenamefont {Kay}, \citenamefont
  {Moya-Cessa}, \citenamefont {Nolte}, \citenamefont {Kwek}, \citenamefont
  {Rodr{\'\i}guez-Lara}, \citenamefont {Szameit},\ and\ \citenamefont
  {Christodoulides}}]{perez2013coherent}%
  \BibitemOpen
  \bibfield  {author} {\bibinfo {author} {\bibfnamefont {A.}~\bibnamefont
  {Perez-Leija}}, \bibinfo {author} {\bibfnamefont {R.}~\bibnamefont {Keil}},
  \bibinfo {author} {\bibfnamefont {A.}~\bibnamefont {Kay}}, \bibinfo {author}
  {\bibfnamefont {H.}~\bibnamefont {Moya-Cessa}}, \bibinfo {author}
  {\bibfnamefont {S.}~\bibnamefont {Nolte}}, \bibinfo {author} {\bibfnamefont
  {L.-C.}\ \bibnamefont {Kwek}}, \bibinfo {author} {\bibfnamefont {B.~M.}\
  \bibnamefont {Rodr{\'\i}guez-Lara}}, \bibinfo {author} {\bibfnamefont
  {A.}~\bibnamefont {Szameit}},\ and\ \bibinfo {author} {\bibfnamefont {D.~N.}\
  \bibnamefont {Christodoulides}},\ }\href@noop {} {\bibfield  {journal}
  {\bibinfo  {journal} {Phys. Rev. A}\ }\textbf {\bibinfo {volume} {87}},\
  \bibinfo {pages} {012309} (\bibinfo {year} {2013})}\BibitemShut {NoStop}%
\bibitem [{\citenamefont {Hahn}\ \emph {et~al.}(2019)\citenamefont {Hahn},
  \citenamefont {Pappa},\ and\ \citenamefont {Eisert}}]{hahn2019quantum}%
  \BibitemOpen
  \bibfield  {author} {\bibinfo {author} {\bibfnamefont {F.}~\bibnamefont
  {Hahn}}, \bibinfo {author} {\bibfnamefont {A.}~\bibnamefont {Pappa}},\ and\
  \bibinfo {author} {\bibfnamefont {J.}~\bibnamefont {Eisert}},\ }\href@noop {}
  {\bibfield  {journal} {\bibinfo  {journal} {npj Quantum Inf.}\ }\textbf
  {\bibinfo {volume} {5}},\ \bibinfo {pages} {1} (\bibinfo {year}
  {2019})}\BibitemShut {NoStop}%
\bibitem [{\citenamefont {Zhan}\ \emph {et~al.}(2014)\citenamefont {Zhan},
  \citenamefont {Qin}, \citenamefont {Bian}, \citenamefont {Li},\ and\
  \citenamefont {Xue}}]{zhan2014perfect}%
  \BibitemOpen
  \bibfield  {author} {\bibinfo {author} {\bibfnamefont {X.}~\bibnamefont
  {Zhan}}, \bibinfo {author} {\bibfnamefont {H.}~\bibnamefont {Qin}}, \bibinfo
  {author} {\bibfnamefont {Z.~H.}\ \bibnamefont {Bian}}, \bibinfo {author}
  {\bibfnamefont {J.}~\bibnamefont {Li}},\ and\ \bibinfo {author}
  {\bibfnamefont {P.}~\bibnamefont {Xue}},\ }\href@noop {} {\bibfield
  {journal} {\bibinfo  {journal} {Phys. Rev. A}\ }\textbf {\bibinfo {volume}
  {90}},\ \bibinfo {pages} {012331} (\bibinfo {year} {2014})}\BibitemShut
  {NoStop}%
\bibitem [{\citenamefont {Sazim}\ \emph {et~al.}(2015)\citenamefont {Sazim},
  \citenamefont {Chiranjeevi}, \citenamefont {Chakrabarty},\ and\ \citenamefont
  {Srinathan}}]{sazim2015retrieving}%
  \BibitemOpen
  \bibfield  {author} {\bibinfo {author} {\bibfnamefont {S.}~\bibnamefont
  {Sazim}}, \bibinfo {author} {\bibfnamefont {V.}~\bibnamefont {Chiranjeevi}},
  \bibinfo {author} {\bibfnamefont {I.}~\bibnamefont {Chakrabarty}},\ and\
  \bibinfo {author} {\bibfnamefont {K.}~\bibnamefont {Srinathan}},\ }\href@noop
  {} {\bibfield  {journal} {\bibinfo  {journal} {Quantum Inf. Process}\
  }\textbf {\bibinfo {volume} {14}},\ \bibinfo {pages} {4651} (\bibinfo {year}
  {2015})}\BibitemShut {NoStop}%
\bibitem [{\citenamefont {Hong}\ \emph {et~al.}(1987)\citenamefont {Hong},
  \citenamefont {Ou},\ and\ \citenamefont {Mandel}}]{hong1987measurement}%
  \BibitemOpen
  \bibfield  {author} {\bibinfo {author} {\bibfnamefont {C.~K.}\ \bibnamefont
  {Hong}}, \bibinfo {author} {\bibfnamefont {Z.~Y.}\ \bibnamefont {Ou}},\ and\
  \bibinfo {author} {\bibfnamefont {L.}~\bibnamefont {Mandel}},\ }\href@noop {}
  {\bibfield  {journal} {\bibinfo  {journal} {Phys. Rev. Lett.}\ }\textbf
  {\bibinfo {volume} {59}},\ \bibinfo {pages} {2044} (\bibinfo {year}
  {1987})}\BibitemShut {NoStop}%
\bibitem [{\citenamefont {Dowling}(2008)}]{dowling2008quantum}%
  \BibitemOpen
  \bibfield  {author} {\bibinfo {author} {\bibfnamefont {J.~P.}\ \bibnamefont
  {Dowling}},\ }\href@noop {} {\bibfield  {journal} {\bibinfo  {journal}
  {Contemp. Phys.}\ }\textbf {\bibinfo {volume} {49}},\ \bibinfo {pages} {125}
  (\bibinfo {year} {2008})}\BibitemShut {NoStop}%
\bibitem [{\citenamefont {Motes}\ \emph {et~al.}(2015)\citenamefont {Motes},
  \citenamefont {Olson}, \citenamefont {Rabeaux}, \citenamefont {Dowling},
  \citenamefont {Olson},\ and\ \citenamefont {Rohde}}]{motes2015linear}%
  \BibitemOpen
  \bibfield  {author} {\bibinfo {author} {\bibfnamefont {K.~R.}\ \bibnamefont
  {Motes}}, \bibinfo {author} {\bibfnamefont {J.~P.}\ \bibnamefont {Olson}},
  \bibinfo {author} {\bibfnamefont {E.~J.}\ \bibnamefont {Rabeaux}}, \bibinfo
  {author} {\bibfnamefont {J.~P.}\ \bibnamefont {Dowling}}, \bibinfo {author}
  {\bibfnamefont {S.~J.}\ \bibnamefont {Olson}},\ and\ \bibinfo {author}
  {\bibfnamefont {P.~P.}\ \bibnamefont {Rohde}},\ }\href@noop {} {\bibfield
  {journal} {\bibinfo  {journal} {Phys. Rev. Lett.}\ }\textbf {\bibinfo
  {volume} {114}},\ \bibinfo {pages} {170802} (\bibinfo {year}
  {2015})}\BibitemShut {NoStop}%
\bibitem [{\citenamefont {Osawa}\ \emph {et~al.}(2020)\citenamefont {Osawa},
  \citenamefont {Simon},\ and\ \citenamefont {Sergienko}}]{osawa2020higher}%
  \BibitemOpen
  \bibfield  {author} {\bibinfo {author} {\bibfnamefont {S.}~\bibnamefont
  {Osawa}}, \bibinfo {author} {\bibfnamefont {D.~S.}\ \bibnamefont {Simon}},\
  and\ \bibinfo {author} {\bibfnamefont {A.~V.}\ \bibnamefont {Sergienko}},\
  }\href@noop {} {\bibfield  {journal} {\bibinfo  {journal} {Phys. Rev. A}\
  }\textbf {\bibinfo {volume} {102}},\ \bibinfo {pages} {063712} (\bibinfo
  {year} {2020})}\BibitemShut {NoStop}%
\bibitem [{\citenamefont {Minder}\ \emph {et~al.}(2019)\citenamefont {Minder},
  \citenamefont {Pittaluga}, \citenamefont {Roberts}, \citenamefont
  {Lucamarini}, \citenamefont {Dynes}, \citenamefont {Yuan},\ and\
  \citenamefont {Shields}}]{minder2019experimental}%
  \BibitemOpen
  \bibfield  {author} {\bibinfo {author} {\bibfnamefont {M.}~\bibnamefont
  {Minder}}, \bibinfo {author} {\bibfnamefont {M.}~\bibnamefont {Pittaluga}},
  \bibinfo {author} {\bibfnamefont {G.}~\bibnamefont {Roberts}}, \bibinfo
  {author} {\bibfnamefont {M.}~\bibnamefont {Lucamarini}}, \bibinfo {author}
  {\bibfnamefont {J.}~\bibnamefont {Dynes}}, \bibinfo {author} {\bibfnamefont
  {Z.}~\bibnamefont {Yuan}},\ and\ \bibinfo {author} {\bibfnamefont
  {A.}~\bibnamefont {Shields}},\ }\href@noop {} {\bibfield  {journal} {\bibinfo
   {journal} {Nat. Photon.}\ }\textbf {\bibinfo {volume} {13}},\ \bibinfo
  {pages} {334} (\bibinfo {year} {2019})}\BibitemShut {NoStop}%
\bibitem [{\citenamefont {Bunandar}\ \emph {et~al.}(2018)\citenamefont
  {Bunandar}, \citenamefont {Lentine}, \citenamefont {Lee}, \citenamefont
  {Cai}, \citenamefont {Long}, \citenamefont {Boynton}, \citenamefont
  {Martinez}, \citenamefont {DeRose}, \citenamefont {Chen}, \citenamefont
  {Grein} \emph {et~al.}}]{bunandar2018metropolitan}%
  \BibitemOpen
  \bibfield  {author} {\bibinfo {author} {\bibfnamefont {D.}~\bibnamefont
  {Bunandar}}, \bibinfo {author} {\bibfnamefont {A.}~\bibnamefont {Lentine}},
  \bibinfo {author} {\bibfnamefont {C.}~\bibnamefont {Lee}}, \bibinfo {author}
  {\bibfnamefont {H.}~\bibnamefont {Cai}}, \bibinfo {author} {\bibfnamefont
  {C.~M.}\ \bibnamefont {Long}}, \bibinfo {author} {\bibfnamefont
  {N.}~\bibnamefont {Boynton}}, \bibinfo {author} {\bibfnamefont
  {N.}~\bibnamefont {Martinez}}, \bibinfo {author} {\bibfnamefont
  {C.}~\bibnamefont {DeRose}}, \bibinfo {author} {\bibfnamefont
  {C.}~\bibnamefont {Chen}}, \bibinfo {author} {\bibfnamefont {M.}~\bibnamefont
  {Grein}}, \emph {et~al.},\ }\href@noop {} {\bibfield  {journal} {\bibinfo
  {journal} {Phys. Rev. X}\ }\textbf {\bibinfo {volume} {8}},\ \bibinfo {pages}
  {021009} (\bibinfo {year} {2018})}\BibitemShut {NoStop}%
\bibitem [{\citenamefont {Bhaskar}\ \emph {et~al.}(2020)\citenamefont
  {Bhaskar}, \citenamefont {Riedinger}, \citenamefont {Machielse},
  \citenamefont {Levonian}, \citenamefont {Nguyen}, \citenamefont {Knall},
  \citenamefont {Park}, \citenamefont {Englund}, \citenamefont {Lon{\v{c}}ar},
  \citenamefont {Sukachev} \emph {et~al.}}]{bhaskar2020experimental}%
  \BibitemOpen
  \bibfield  {author} {\bibinfo {author} {\bibfnamefont {M.~K.}\ \bibnamefont
  {Bhaskar}}, \bibinfo {author} {\bibfnamefont {R.}~\bibnamefont {Riedinger}},
  \bibinfo {author} {\bibfnamefont {B.}~\bibnamefont {Machielse}}, \bibinfo
  {author} {\bibfnamefont {D.~S.}\ \bibnamefont {Levonian}}, \bibinfo {author}
  {\bibfnamefont {C.~T.}\ \bibnamefont {Nguyen}}, \bibinfo {author}
  {\bibfnamefont {E.~N.}\ \bibnamefont {Knall}}, \bibinfo {author}
  {\bibfnamefont {H.}~\bibnamefont {Park}}, \bibinfo {author} {\bibfnamefont
  {D.}~\bibnamefont {Englund}}, \bibinfo {author} {\bibfnamefont
  {M.}~\bibnamefont {Lon{\v{c}}ar}}, \bibinfo {author} {\bibfnamefont {D.~D.}\
  \bibnamefont {Sukachev}}, \emph {et~al.},\ }\href@noop {} {\bibfield
  {journal} {\bibinfo  {journal} {Nature}\ }\textbf {\bibinfo {volume} {580}},\
  \bibinfo {pages} {60} (\bibinfo {year} {2020})}\BibitemShut {NoStop}%
\bibitem [{\citenamefont {Larsen}\ \emph {et~al.}(2019)\citenamefont {Larsen},
  \citenamefont {Guo}, \citenamefont {Breum}, \citenamefont
  {Neergaard-Nielsen},\ and\ \citenamefont
  {Andersen}}]{larsen2019deterministic}%
  \BibitemOpen
  \bibfield  {author} {\bibinfo {author} {\bibfnamefont {M.~V.}\ \bibnamefont
  {Larsen}}, \bibinfo {author} {\bibfnamefont {X.}~\bibnamefont {Guo}},
  \bibinfo {author} {\bibfnamefont {C.~R.}\ \bibnamefont {Breum}}, \bibinfo
  {author} {\bibfnamefont {J.~S.}\ \bibnamefont {Neergaard-Nielsen}},\ and\
  \bibinfo {author} {\bibfnamefont {U.~L.}\ \bibnamefont {Andersen}},\
  }\href@noop {} {\bibfield  {journal} {\bibinfo  {journal} {Science}\ }\textbf
  {\bibinfo {volume} {366}},\ \bibinfo {pages} {369} (\bibinfo {year}
  {2019})}\BibitemShut {NoStop}%
\bibitem [{\citenamefont {Asavanant}\ \emph {et~al.}(2019)\citenamefont
  {Asavanant}, \citenamefont {Shiozawa}, \citenamefont {Yokoyama},
  \citenamefont {Charoensombutamon}, \citenamefont {Emura}, \citenamefont
  {Alexander}, \citenamefont {Takeda}, \citenamefont {Yoshikawa}, \citenamefont
  {Menicucci}, \citenamefont {Yonezawa} \emph
  {et~al.}}]{asavanant2019generation}%
  \BibitemOpen
  \bibfield  {author} {\bibinfo {author} {\bibfnamefont {W.}~\bibnamefont
  {Asavanant}}, \bibinfo {author} {\bibfnamefont {Y.}~\bibnamefont {Shiozawa}},
  \bibinfo {author} {\bibfnamefont {S.}~\bibnamefont {Yokoyama}}, \bibinfo
  {author} {\bibfnamefont {B.}~\bibnamefont {Charoensombutamon}}, \bibinfo
  {author} {\bibfnamefont {H.}~\bibnamefont {Emura}}, \bibinfo {author}
  {\bibfnamefont {R.~N.}\ \bibnamefont {Alexander}}, \bibinfo {author}
  {\bibfnamefont {S.}~\bibnamefont {Takeda}}, \bibinfo {author} {\bibfnamefont
  {J.-i.}\ \bibnamefont {Yoshikawa}}, \bibinfo {author} {\bibfnamefont {N.~C.}\
  \bibnamefont {Menicucci}}, \bibinfo {author} {\bibfnamefont {H.}~\bibnamefont
  {Yonezawa}}, \emph {et~al.},\ }\href@noop {} {\bibfield  {journal} {\bibinfo
  {journal} {Science}\ }\textbf {\bibinfo {volume} {366}},\ \bibinfo {pages}
  {373} (\bibinfo {year} {2019})}\BibitemShut {NoStop}%
\bibitem [{\citenamefont {Wang}\ \emph {et~al.}(2017)\citenamefont {Wang},
  \citenamefont {Paesani}, \citenamefont {Santagati}, \citenamefont {Knauer},
  \citenamefont {Gentile}, \citenamefont {Wiebe}, \citenamefont {Petruzzella},
  \citenamefont {O’Brien}, \citenamefont {Rarity}, \citenamefont {Laing}
  \emph {et~al.}}]{wang2017experimental}%
  \BibitemOpen
  \bibfield  {author} {\bibinfo {author} {\bibfnamefont {J.}~\bibnamefont
  {Wang}}, \bibinfo {author} {\bibfnamefont {S.}~\bibnamefont {Paesani}},
  \bibinfo {author} {\bibfnamefont {R.}~\bibnamefont {Santagati}}, \bibinfo
  {author} {\bibfnamefont {S.}~\bibnamefont {Knauer}}, \bibinfo {author}
  {\bibfnamefont {A.~A.}\ \bibnamefont {Gentile}}, \bibinfo {author}
  {\bibfnamefont {N.}~\bibnamefont {Wiebe}}, \bibinfo {author} {\bibfnamefont
  {M.}~\bibnamefont {Petruzzella}}, \bibinfo {author} {\bibfnamefont {J.~L.}\
  \bibnamefont {O’Brien}}, \bibinfo {author} {\bibfnamefont {J.~G.}\
  \bibnamefont {Rarity}}, \bibinfo {author} {\bibfnamefont {A.}~\bibnamefont
  {Laing}}, \emph {et~al.},\ }\href@noop {} {\bibfield  {journal} {\bibinfo
  {journal} {Nat. Phys.}\ }\textbf {\bibinfo {volume} {13}},\ \bibinfo {pages}
  {551} (\bibinfo {year} {2017})}\BibitemShut {NoStop}%
\bibitem [{\citenamefont {Peruzzo}\ \emph {et~al.}(2014)\citenamefont
  {Peruzzo}, \citenamefont {McClean}, \citenamefont {Shadbolt}, \citenamefont
  {Yung}, \citenamefont {Zhou}, \citenamefont {Love}, \citenamefont
  {Aspuru-Guzik},\ and\ \citenamefont {O’brien}}]{peruzzo2014variational}%
  \BibitemOpen
  \bibfield  {author} {\bibinfo {author} {\bibfnamefont {A.}~\bibnamefont
  {Peruzzo}}, \bibinfo {author} {\bibfnamefont {J.}~\bibnamefont {McClean}},
  \bibinfo {author} {\bibfnamefont {P.}~\bibnamefont {Shadbolt}}, \bibinfo
  {author} {\bibfnamefont {M.-H.}\ \bibnamefont {Yung}}, \bibinfo {author}
  {\bibfnamefont {X.-Q.}\ \bibnamefont {Zhou}}, \bibinfo {author}
  {\bibfnamefont {P.~J.}\ \bibnamefont {Love}}, \bibinfo {author}
  {\bibfnamefont {A.}~\bibnamefont {Aspuru-Guzik}},\ and\ \bibinfo {author}
  {\bibfnamefont {J.~L.}\ \bibnamefont {O’brien}},\ }\href@noop {} {\bibfield
   {journal} {\bibinfo  {journal} {Nat. Commun.}\ }\textbf {\bibinfo {volume}
  {5}},\ \bibinfo {pages} {1} (\bibinfo {year} {2014})}\BibitemShut {NoStop}%
\bibitem [{\citenamefont {Van~Campenhout}\ \emph {et~al.}(2010)\citenamefont
  {Van~Campenhout}, \citenamefont {Green}, \citenamefont {Assefa},\ and\
  \citenamefont {Vlasov}}]{van2010integrated}%
  \BibitemOpen
  \bibfield  {author} {\bibinfo {author} {\bibfnamefont {J.}~\bibnamefont
  {Van~Campenhout}}, \bibinfo {author} {\bibfnamefont {W.~M.}\ \bibnamefont
  {Green}}, \bibinfo {author} {\bibfnamefont {S.}~\bibnamefont {Assefa}},\ and\
  \bibinfo {author} {\bibfnamefont {Y.~A.}\ \bibnamefont {Vlasov}},\
  }\href@noop {} {\bibfield  {journal} {\bibinfo  {journal} {Opt. Lett.}\
  }\textbf {\bibinfo {volume} {35}},\ \bibinfo {pages} {1013} (\bibinfo {year}
  {2010})}\BibitemShut {NoStop}%
\bibitem [{\citenamefont {Masood}\ \emph {et~al.}(2013)\citenamefont {Masood},
  \citenamefont {Pantouvaki}, \citenamefont {Lepage}, \citenamefont {Verheyen},
  \citenamefont {Van~Campenhout}, \citenamefont {Absil}, \citenamefont
  {Van~Thourhout},\ and\ \citenamefont {Bogaerts}}]{masood2013comparison}%
  \BibitemOpen
  \bibfield  {author} {\bibinfo {author} {\bibfnamefont {A.}~\bibnamefont
  {Masood}}, \bibinfo {author} {\bibfnamefont {M.}~\bibnamefont {Pantouvaki}},
  \bibinfo {author} {\bibfnamefont {G.}~\bibnamefont {Lepage}}, \bibinfo
  {author} {\bibfnamefont {P.}~\bibnamefont {Verheyen}}, \bibinfo {author}
  {\bibfnamefont {J.}~\bibnamefont {Van~Campenhout}}, \bibinfo {author}
  {\bibfnamefont {P.}~\bibnamefont {Absil}}, \bibinfo {author} {\bibfnamefont
  {D.}~\bibnamefont {Van~Thourhout}},\ and\ \bibinfo {author} {\bibfnamefont
  {W.}~\bibnamefont {Bogaerts}},\ }in\ \href@noop {} {\emph {\bibinfo
  {booktitle} {10th International Conference on Group IV Photonics}}}\
  (\bibinfo {organization} {IEEE},\ \bibinfo {year} {2013})\ pp.\ \bibinfo
  {pages} {83--84}\BibitemShut {NoStop}%
\bibitem [{\citenamefont {Wang}\ \emph {et~al.}(2019)\citenamefont {Wang},
  \citenamefont {Qin}, \citenamefont {Ding}, \citenamefont {Chen},
  \citenamefont {Chen}, \citenamefont {You}, \citenamefont {He}, \citenamefont
  {Jiang}, \citenamefont {You}, \citenamefont {Wang} \emph
  {et~al.}}]{wang2019boson}%
  \BibitemOpen
  \bibfield  {author} {\bibinfo {author} {\bibfnamefont {H.}~\bibnamefont
  {Wang}}, \bibinfo {author} {\bibfnamefont {J.}~\bibnamefont {Qin}}, \bibinfo
  {author} {\bibfnamefont {X.}~\bibnamefont {Ding}}, \bibinfo {author}
  {\bibfnamefont {M.-C.}\ \bibnamefont {Chen}}, \bibinfo {author}
  {\bibfnamefont {S.}~\bibnamefont {Chen}}, \bibinfo {author} {\bibfnamefont
  {X.}~\bibnamefont {You}}, \bibinfo {author} {\bibfnamefont {Y.-M.}\
  \bibnamefont {He}}, \bibinfo {author} {\bibfnamefont {X.}~\bibnamefont
  {Jiang}}, \bibinfo {author} {\bibfnamefont {L.}~\bibnamefont {You}}, \bibinfo
  {author} {\bibfnamefont {Z.}~\bibnamefont {Wang}}, \emph {et~al.},\
  }\href@noop {} {\bibfield  {journal} {\bibinfo  {journal} {Phys. Rev. Lett.}\
  }\textbf {\bibinfo {volume} {123}},\ \bibinfo {pages} {250503} (\bibinfo
  {year} {2019})}\BibitemShut {NoStop}%
\bibitem [{\citenamefont {Tillmann}\ \emph {et~al.}(2013)\citenamefont
  {Tillmann}, \citenamefont {Daki{\'c}}, \citenamefont {Heilmann},
  \citenamefont {Nolte}, \citenamefont {Szameit},\ and\ \citenamefont
  {Walther}}]{tillmann2013experimental}%
  \BibitemOpen
  \bibfield  {author} {\bibinfo {author} {\bibfnamefont {M.}~\bibnamefont
  {Tillmann}}, \bibinfo {author} {\bibfnamefont {B.}~\bibnamefont {Daki{\'c}}},
  \bibinfo {author} {\bibfnamefont {R.}~\bibnamefont {Heilmann}}, \bibinfo
  {author} {\bibfnamefont {S.}~\bibnamefont {Nolte}}, \bibinfo {author}
  {\bibfnamefont {A.}~\bibnamefont {Szameit}},\ and\ \bibinfo {author}
  {\bibfnamefont {P.}~\bibnamefont {Walther}},\ }\href@noop {} {\bibfield
  {journal} {\bibinfo  {journal} {Nat. Photon.}\ }\textbf {\bibinfo {volume}
  {7}},\ \bibinfo {pages} {540} (\bibinfo {year} {2013})}\BibitemShut {NoStop}%
\bibitem [{\citenamefont {Bentivegna}\ \emph {et~al.}(2015)\citenamefont
  {Bentivegna}, \citenamefont {Spagnolo}, \citenamefont {Vitelli},
  \citenamefont {Flamini}, \citenamefont {Viggianiello}, \citenamefont
  {Latmiral}, \citenamefont {Mataloni}, \citenamefont {Brod}, \citenamefont
  {Galv{\~a}o}, \citenamefont {Crespi} \emph
  {et~al.}}]{bentivegna2015experimental}%
  \BibitemOpen
  \bibfield  {author} {\bibinfo {author} {\bibfnamefont {M.}~\bibnamefont
  {Bentivegna}}, \bibinfo {author} {\bibfnamefont {N.}~\bibnamefont
  {Spagnolo}}, \bibinfo {author} {\bibfnamefont {C.}~\bibnamefont {Vitelli}},
  \bibinfo {author} {\bibfnamefont {F.}~\bibnamefont {Flamini}}, \bibinfo
  {author} {\bibfnamefont {N.}~\bibnamefont {Viggianiello}}, \bibinfo {author}
  {\bibfnamefont {L.}~\bibnamefont {Latmiral}}, \bibinfo {author}
  {\bibfnamefont {P.}~\bibnamefont {Mataloni}}, \bibinfo {author}
  {\bibfnamefont {D.~J.}\ \bibnamefont {Brod}}, \bibinfo {author}
  {\bibfnamefont {E.~F.}\ \bibnamefont {Galv{\~a}o}}, \bibinfo {author}
  {\bibfnamefont {A.}~\bibnamefont {Crespi}}, \emph {et~al.},\ }\href@noop {}
  {\bibfield  {journal} {\bibinfo  {journal} {Sci. Adv.}\ }\textbf {\bibinfo
  {volume} {1}},\ \bibinfo {pages} {e1400255} (\bibinfo {year}
  {2015})}\BibitemShut {NoStop}%
\bibitem [{\citenamefont {Simon}\ \emph {et~al.}(2020)\citenamefont {Simon},
  \citenamefont {Osawa},\ and\ \citenamefont {Sergienko}}]{simon2020quantum}%
  \BibitemOpen
  \bibfield  {author} {\bibinfo {author} {\bibfnamefont {D.~S.}\ \bibnamefont
  {Simon}}, \bibinfo {author} {\bibfnamefont {S.}~\bibnamefont {Osawa}},\ and\
  \bibinfo {author} {\bibfnamefont {A.~V.}\ \bibnamefont {Sergienko}},\
  }\href@noop {} {\bibfield  {journal} {\bibinfo  {journal} {Phys. Rev. A}\
  }\textbf {\bibinfo {volume} {101}},\ \bibinfo {pages} {032118} (\bibinfo
  {year} {2020})}\BibitemShut {NoStop}%
\bibitem [{\citenamefont {Osawa}\ \emph {et~al.}(2018)\citenamefont {Osawa},
  \citenamefont {Simon},\ and\ \citenamefont
  {Sergienko}}]{osawa2018experimental}%
  \BibitemOpen
  \bibfield  {author} {\bibinfo {author} {\bibfnamefont {S.}~\bibnamefont
  {Osawa}}, \bibinfo {author} {\bibfnamefont {D.~S.}\ \bibnamefont {Simon}},\
  and\ \bibinfo {author} {\bibfnamefont {A.~V.}\ \bibnamefont {Sergienko}},\
  }\href@noop {} {\bibfield  {journal} {\bibinfo  {journal} {Opt. Express}\
  }\textbf {\bibinfo {volume} {26}},\ \bibinfo {pages} {27201} (\bibinfo {year}
  {2018})}\BibitemShut {NoStop}%
\bibitem [{\citenamefont {Kim}\ \emph {et~al.}(2021)\citenamefont {Kim},
  \citenamefont {Lee}, \citenamefont {Hong}, \citenamefont {Cho}, \citenamefont
  {Lee}, \citenamefont {Kim},\ and\ \citenamefont
  {Lim}}]{kim2021implementation}%
  \BibitemOpen
  \bibfield  {author} {\bibinfo {author} {\bibfnamefont {I.}~\bibnamefont
  {Kim}}, \bibinfo {author} {\bibfnamefont {D.}~\bibnamefont {Lee}}, \bibinfo
  {author} {\bibfnamefont {S.}~\bibnamefont {Hong}}, \bibinfo {author}
  {\bibfnamefont {Y.-W.}\ \bibnamefont {Cho}}, \bibinfo {author} {\bibfnamefont
  {K.~J.}\ \bibnamefont {Lee}}, \bibinfo {author} {\bibfnamefont {Y.-S.}\
  \bibnamefont {Kim}},\ and\ \bibinfo {author} {\bibfnamefont {H.-T.}\
  \bibnamefont {Lim}},\ }\href@noop {} {\bibfield  {journal} {\bibinfo
  {journal} {arXiv preprint arXiv:2106.13473}\ } (\bibinfo {year}
  {2021})}\BibitemShut {NoStop}%
\bibitem [{\citenamefont {Hillery}\ \emph {et~al.}(1999)\citenamefont
  {Hillery}, \citenamefont {Bu{\v{z}}ek},\ and\ \citenamefont
  {Berthiaume}}]{hillery1999quantum}%
  \BibitemOpen
  \bibfield  {author} {\bibinfo {author} {\bibfnamefont {M.}~\bibnamefont
  {Hillery}}, \bibinfo {author} {\bibfnamefont {V.}~\bibnamefont
  {Bu{\v{z}}ek}},\ and\ \bibinfo {author} {\bibfnamefont {A.}~\bibnamefont
  {Berthiaume}},\ }\href@noop {} {\bibfield  {journal} {\bibinfo  {journal}
  {Phys. Rev. A}\ }\textbf {\bibinfo {volume} {59}},\ \bibinfo {pages} {1829}
  (\bibinfo {year} {1999})}\BibitemShut {NoStop}%
\bibitem [{\citenamefont {Tittel}\ \emph {et~al.}(2001)\citenamefont {Tittel},
  \citenamefont {Zbinden},\ and\ \citenamefont
  {Gisin}}]{tittel2001experimental}%
  \BibitemOpen
  \bibfield  {author} {\bibinfo {author} {\bibfnamefont {W.}~\bibnamefont
  {Tittel}}, \bibinfo {author} {\bibfnamefont {H.}~\bibnamefont {Zbinden}},\
  and\ \bibinfo {author} {\bibfnamefont {N.}~\bibnamefont {Gisin}},\
  }\href@noop {} {\bibfield  {journal} {\bibinfo  {journal} {Phys. Rev. A}\
  }\textbf {\bibinfo {volume} {63}},\ \bibinfo {pages} {042301} (\bibinfo
  {year} {2001})}\BibitemShut {NoStop}%
\bibitem [{\citenamefont {Bell}\ \emph {et~al.}(2014)\citenamefont {Bell},
  \citenamefont {Markham}, \citenamefont {Herrera-Mart{\'\i}}, \citenamefont
  {Marin}, \citenamefont {Wadsworth}, \citenamefont {Rarity},\ and\
  \citenamefont {Tame}}]{bell2014experimental}%
  \BibitemOpen
  \bibfield  {author} {\bibinfo {author} {\bibfnamefont {B.}~\bibnamefont
  {Bell}}, \bibinfo {author} {\bibfnamefont {D.}~\bibnamefont {Markham}},
  \bibinfo {author} {\bibfnamefont {D.}~\bibnamefont {Herrera-Mart{\'\i}}},
  \bibinfo {author} {\bibfnamefont {A.}~\bibnamefont {Marin}}, \bibinfo
  {author} {\bibfnamefont {W.}~\bibnamefont {Wadsworth}}, \bibinfo {author}
  {\bibfnamefont {J.}~\bibnamefont {Rarity}},\ and\ \bibinfo {author}
  {\bibfnamefont {M.}~\bibnamefont {Tame}},\ }\href@noop {} {\bibfield
  {journal} {\bibinfo  {journal} {Nat. Commun.}\ }\textbf {\bibinfo {volume}
  {5}},\ \bibinfo {pages} {3658} (\bibinfo {year} {2014})}\BibitemShut
  {NoStop}%
\bibitem [{\citenamefont {Williams}\ \emph {et~al.}(2019)\citenamefont
  {Williams}, \citenamefont {Lukens}, \citenamefont {Peters}, \citenamefont
  {Qi},\ and\ \citenamefont {Grice}}]{williams2019quantum}%
  \BibitemOpen
  \bibfield  {author} {\bibinfo {author} {\bibfnamefont {B.~P.}\ \bibnamefont
  {Williams}}, \bibinfo {author} {\bibfnamefont {J.~M.}\ \bibnamefont
  {Lukens}}, \bibinfo {author} {\bibfnamefont {N.~A.}\ \bibnamefont {Peters}},
  \bibinfo {author} {\bibfnamefont {B.}~\bibnamefont {Qi}},\ and\ \bibinfo
  {author} {\bibfnamefont {W.~P.}\ \bibnamefont {Grice}},\ }\href@noop {}
  {\bibfield  {journal} {\bibinfo  {journal} {Phys. Rev. A}\ }\textbf {\bibinfo
  {volume} {99}},\ \bibinfo {pages} {062311} (\bibinfo {year}
  {2019})}\BibitemShut {NoStop}%
\bibitem [{\citenamefont {Lidar}\ \emph {et~al.}(1998)\citenamefont {Lidar},
  \citenamefont {Chuang},\ and\ \citenamefont {Whaley}}]{lidar1998decoherence}%
  \BibitemOpen
  \bibfield  {author} {\bibinfo {author} {\bibfnamefont {D.~A.}\ \bibnamefont
  {Lidar}}, \bibinfo {author} {\bibfnamefont {I.~L.}\ \bibnamefont {Chuang}},\
  and\ \bibinfo {author} {\bibfnamefont {K.~B.}\ \bibnamefont {Whaley}},\
  }\href@noop {} {\bibfield  {journal} {\bibinfo  {journal} {Physical Review
  Letters}\ }\textbf {\bibinfo {volume} {81}},\ \bibinfo {pages} {2594}
  (\bibinfo {year} {1998})}\BibitemShut {NoStop}%
\bibitem [{\citenamefont {Walton}\ \emph {et~al.}(2003)\citenamefont {Walton},
  \citenamefont {Abouraddy}, \citenamefont {Sergienko}, \citenamefont {Saleh},\
  and\ \citenamefont {Teich}}]{walton2003decoherence}%
  \BibitemOpen
  \bibfield  {author} {\bibinfo {author} {\bibfnamefont {Z.~D.}\ \bibnamefont
  {Walton}}, \bibinfo {author} {\bibfnamefont {A.~F.}\ \bibnamefont
  {Abouraddy}}, \bibinfo {author} {\bibfnamefont {A.~V.}\ \bibnamefont
  {Sergienko}}, \bibinfo {author} {\bibfnamefont {B.~E.}\ \bibnamefont
  {Saleh}},\ and\ \bibinfo {author} {\bibfnamefont {M.~C.}\ \bibnamefont
  {Teich}},\ }\href@noop {} {\bibfield  {journal} {\bibinfo  {journal}
  {Physical Review Letters}\ }\textbf {\bibinfo {volume} {91}},\ \bibinfo
  {pages} {087901} (\bibinfo {year} {2003})}\BibitemShut {NoStop}%
\end{thebibliography}%

\end{document}